\documentclass[floatfix,prd,epsfig,nofootinbib,superscriptaddress,onecolumn,amssymb]{revtex4}

\usepackage{slashed}
\usepackage{slashed}
\usepackage{graphicx,color}
\usepackage{epsfig}
\usepackage{subfigure}
\usepackage{epsfig}
\usepackage{dcolumn}
\usepackage{bm}
\usepackage{color}

\def\lsim{\mathrel{\rlap{\lower4pt\hbox{\hskip1pt$\sim$}}
    \raise1pt\hbox{$<$}}}         
\def\gsim{\mathrel{\rlap{\lower4pt\hbox{\hskip1pt$\sim$}}
    \raise1pt\hbox{$>$}}}         
    
    \newcommand{\nc}{\newcommand}  

\nc{\beq}{\begin{equation}}  
\nc{\eeq}{\end{equation}}  
\nc{\beqa}{\begin{eqnarray}}  
\nc{\eeqa}{\end{eqnarray}}  
\nc{\bea}{\begin{eqnarray}}  
\nc{\eea}{\end{eqnarray}}  
\nc{\ra}{\rightarrow}  
\nc{\slsh}{\slash\hspace*{-0.22cm}}

\def\Re{{\cal R \mskip-4mu \lower.1ex \hbox{\it e}\,}}
\def\Im{{\cal I \mskip-5mu \lower.1ex \hbox{\it m}\,}}
\def\be{\begin{equation}}
\def\ee{\end{equation}}
\def\bea{\begin{eqnarray}}
\def\eea{\end{eqnarray}}
\def\bit{\begin{itemize}}
\def\eit{\end{itemize}}
\nc{\eref}[1]{(\ref{#1})}
\nc{\Eref}[1]{Eq.~(\ref{#1})}

\nc{\vev}[1]{ \left\langle {#1} \right\rangle }
\nc{\bra}[1]{ \langle {#1} | }
\nc{\ket}[1]{ | {#1} \rangle }
\nc{\fb}{\,{\rm fb}^{-1}}
\nc{\ev}{{\rm eV}}
\nc{\kev}{{\rm keV}}
\nc{\Mev}{{\rm MeV}}
\nc{\gev}{{\rm GeV}}
\nc{\tev}{{\rm TeV}}
\nc{\mev}{{\rm MeV}}

\def\ee{e^+e^-}

\def\msb{{\bar{\ssstyle M \kern -1pt S}}}

\begin{document}

\title{The MOLLER Experiment: An Ultra-Precise Measurement of the Weak Mixing Angle using M{\o}ller Scattering}

\author{The MOLLER Collaboration\footnote{Corresponding Author: Krishna S. Kumar}}\email{krishna.kumar@stonybrook.edu}
\noaffiliation


\vspace{3cm}

\begin{abstract}
The
physics case and an experimental overview of the MOLLER ({\bf M}easurement {\bf O}f a {\bf L}epton {\bf L}epton {\bf E}lectroweak {\bf R}eaction) experiment at the 12 GeV upgraded Jefferson Lab
are presented.  
A highlight of the Fundamental Symmetries subfield of the 2007 NSAC Long Range Plan was the SLAC E158 measurement of the parity-violating asymmetry  $A_{PV}$ in polarized electron-electron (M{\o}ller) scattering.  The proposed MOLLER experiment will improve on this result by a factor of five, yielding the most precise measurement of the weak mixing angle at low or high energy anticipated over the next decade.  This new result would be sensitive to the interference of the electromagnetic amplitude with new neutral current amplitudes as weak as
$\sim 10^{-3}\cdot G_F$ from as yet undiscovered dynamics beyond the Standard Model. The resulting discovery reach is unmatched  by any proposed experiment measuring a flavor- and CP-conserving process  over the next decade, and yields a unique window to new physics at MeV and multi-TeV scales, complementary to direct searches at high energy colliders such as the Large Hadron Collider (LHC).   The experiment takes advantage of the unique opportunity provided by the upgraded electron beam energy, luminosity, and stability at Jefferson Laboratory and the extensive experience accumulated in the community after a round of recent successfully completed parity-violating electron scattering experiments.
\end{abstract}

\maketitle
\section{Executive Summary}

An important strategy to determine the full extent of validity of the electroweak theory and search for new dynamics from MeV to multi-TeV scales
involves indirect probes, where ultra-precise measurements of electroweak observables at energy scales well below the scale of electroweak symmetry breaking are compared to accurate theoretical predictions. The MOLLER project~\cite{MOLLER1,MOLLER2}, like many of the low energy experimental measurements that have been proposed in the Fundamental Symmetries area, pursues 
such a strategy. 
MOLLER proposes
to measure the parity-violating asymmetry in the scattering of longitudinally polarized electrons off unpolarized electrons, using the upgraded 11~GeV beam in Hall~A at Jefferson Laboratory (JLab), to an overall fractional accuracy of 2.4\%. Such a measurement would constitute more than a factor of five improvement in fractional precision over the only other
measurement of the same quantity by the E158 experiment at SLAC~\cite{Anthony:2005pm}. 

The electron beam energy, luminosity and stability at Jefferson Laboratory are uniquely suited to carry out such a measurement. 
The 11 GeV JLab beam at the upgraded facility provides a compelling new opportunity  to achieve a new benchmark in sensitivity. The physics motivation has two important aspects: 
\begin{enumerate}
\item New neutral current interactions are best parameterized model-independently at low energies by effective four-fermion interactions via the quantity $\Lambda/g$, where $g$ characterizes the strength and $\Lambda$ is the scale of the new dynamics. 
The proposed $A_{PV}$ measurement is sensitive to interaction amplitudes as small as $1.5\times 10^{-3}$ times the Fermi constant, $G_F$, which corresponds to a sensitivity of $\Lambda/g =  7.5$~TeV. This would be {\it the} most sensitive probe of new flavor and CP-conserving 
neutral current interactions in the leptonic sector until the advent of a linear collider or a neutrino factory.
Such a measurement
has $\pm 10\ \sigma$ discovery potential in the 
discovery space
allowed by the existing most precise low energy measurements.
\item Within the Standard Model, weak neutral current amplitudes are functions of the weak mixing angle $\sin^2\theta_W$. The two most precise independent determinations of $\sin^2\theta_W$
differ by 3$\sigma$. The world average is consistent with the theoretical prediction for the weak mixing angle assuming the 126 GeV scalar resonance observed at the LHC is the Standard Model (SM) Higgs boson.  However,  choosing one or the other central value ruins this consistency and implies very different new high-energy dynamics. 
The proposed $A_{PV}$ measurement, which would achieve a sensitivity of 
$\delta(\sin^2\theta_W) = \pm 0.00028$, has the same level of precision and interpretability: the best among projected
sensitivies for new measurements at low $Q^2$ or colliders over the next decade.
\end{enumerate}

$A_{PV}$ in M\o ller scattering measures the weak charge of the electron
$Q^e_W$, which is proportional to the product of the electron's vector and axial-vector couplings to the $Z^0$ boson. 
The electroweak theory prediction at tree level in terms of the weak mixing angle is $Q^e_W = 1 - 4\sin^2\theta_W$; this is modified at the 1-loop level~\cite{Czarnecki:1995fw, Czarnecki:2000ic, Erler:2004in} and becomes dependent on the energy scale at which the measurement is carried out, {\em i.e.} $\sin^2\theta_W$ ``runs". 
The prediction for $A_{PV}$ for the proposed experimental design is 
$\approx 33$~parts per billion (ppb) and 
the
goal is to measure this quantity with an overall precision of 0.7 ppb
and thus achieve a 2.4\%\ measurement of $Q^e_W$. Under the assumption of a SM Higgs boson mass of 126 GeV, the theoretical prediction for the MOLLER $A_{PV}$ is known to better than 0.2 ppb accuracy.  The purely leptonic M{\o}ller PV asymmetry is a rare low energy observable whose theoretical uncertainties, especially due to hadronic effects, are well under control.  

The MOLLER experiment would measure a unique observable and be among the most sensitive in terms of discovery reach for flavor- and CP- conserving scattering amplitudes in the next decade; see recent reviews that situate the measurement in broader contexts~\cite{kkreview, Cirigliano:2013lpa, Erler:2013xha}.  It is very complementary to other precision low energy experiments and the energy frontier efforts at the LHC.  If the LHC continues to agree with the Standard Model with high luminosity running at the full 14 TeV energy, then MOLLER will be a significant component of a global strategy to discover signatures of a variety of physics that could escape LHC detection.  Examples include hidden weak scale scenarios such as compressed supersymmetry~\cite{Chao:2014}, lepton number violating amplitudes such as those mediated by doubly charged scalars~\cite{Cirigliano:2004mv}, and light MeV-scale dark matter mediators such as the ``dark'' Z~\cite{Davoudiasl:2012ag,Davoudiasl:2012qa}.  If the LHC observes an anomaly, then MOLLER will have the sensitivity to be part of a few select measurements that will provide important constraints to choose among possible beyond the Standard Model (BSM) scenarios to explain the anomaly.  Examples of such BSM scenarios that have been explicitly considered for MOLLER include: new particles predicted by the Minimal Supersymmetric Standard Model observed through radiative loop effects (R-parity conserving) or tree-level interactions (R-parity violating)~\cite{Kurylov:2003zh,RamseyMusolf:2006vr} and TeV-scale $Z^\prime$s~\cite{Erler:2011iw} which 
arise in many BSM theories.

The 2007 NSAC long range plan report~\cite{nsaclrp07} comprehensively described the opportunities presented by new sensitive indirect probes such as MOLLER, and how they fit into the subfield of Fundamental Symmetries.
One of the overarching questions that serves to define this subfield is: ``What are the unseen forces that were present at the dawn of the universe but disappeared from view as the universe evolved?". To address this question and as part of the third principal recommendation, significant new investments, including MOLLER, were advocated.
Since then,  MOLLER received the highest rating from the JLab Program Advisory Committee (PAC) in January 2009.
In January 2010, 
JLab management organized a Director's review of the experiment chaired by Charles Prescott~\cite{prescottreview}. The committee
gave strong endorsement to the experiment and encouraged the collaboration and the laboratory to develop
a full proposal to obtain construction funding.  
In January 2011, 
the PAC allocated MOLLER's full beamtime request of 344 PAC days. The 2012 NSAC subpanel on the implementation of the Long Range Plan (the Tribble 
 Subcommittee)~\cite{nsacsubcomm12} strongly endorsed the MOLLER project as part of the suite of 
 investments advocated for the subfield of Fundamental Symmetries.
In September 2014,  JLab submitted a 
document describing the science case and the experimental design
to DoE on behalf of the MOLLER collaboration\footnote{See Appendix A for full MOLLER collaboration list.} ($\sim120$ collaborators from 30 institutions representing 6 countries).
Most recently, in September 2014, the experiment underwent a Science Review conducted by the DOE Office of Nuclear Physics.
The goal is to obtain construction funding in fiscal year 2017, with 
the intention of installing the apparatus in fiscal year 2019 and commissioning 
and taking first physics data by 2020.

\section{Physics Motivation}

\begin{figure}[htp]
\noindent
\begin{minipage}{16.cm}
\vspace*{5mm}
\hspace*{15mm}
\begin{center}
\begin{tabular}{cccc}
\includegraphics[width=3.75cm, trim = 20mm 140mm 20mm 140mm]{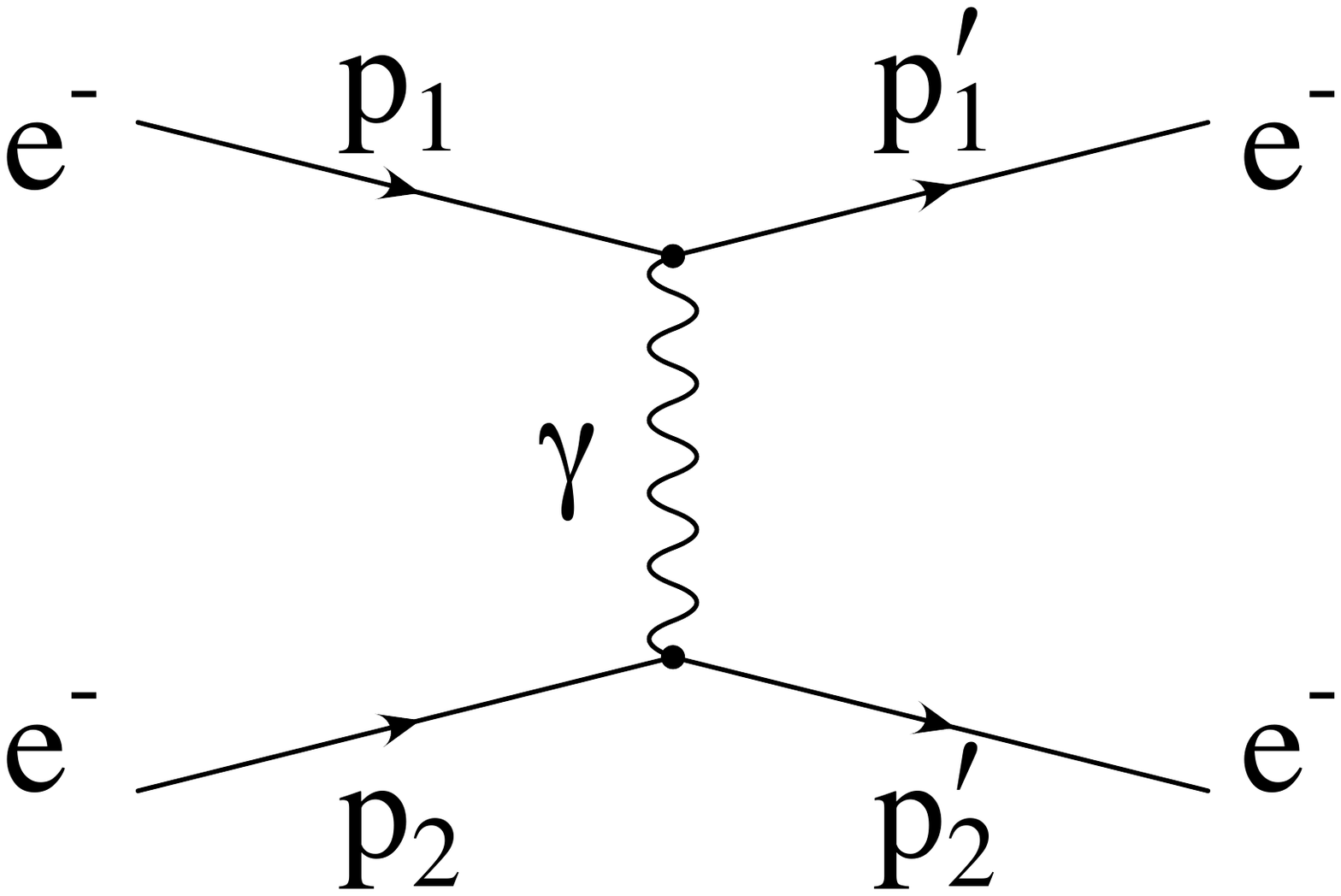}&
\includegraphics[width=3.75cm, trim = 20mm 140mm 20mm 140mm]{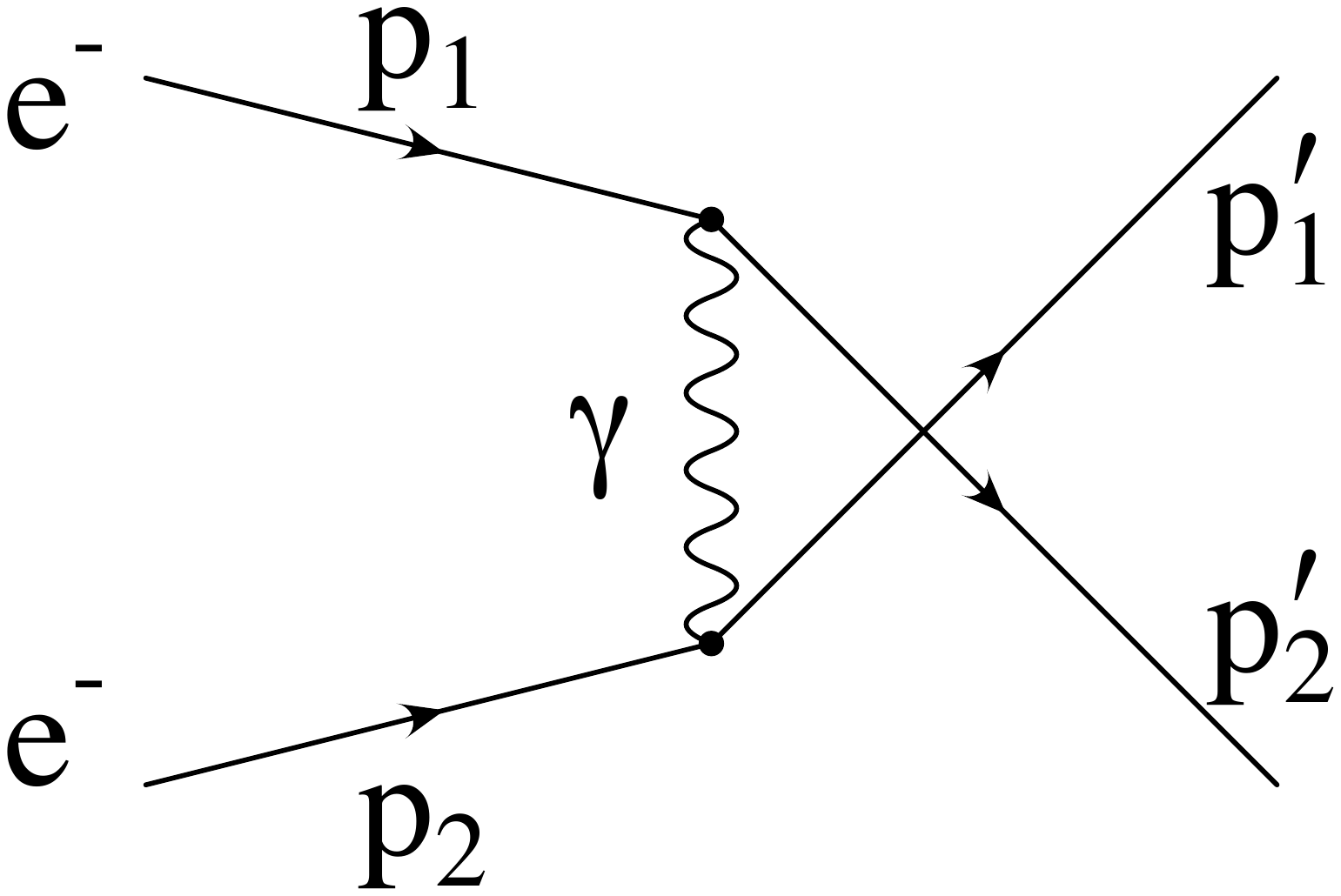}&
\includegraphics[width=3.75cm, trim = 20mm 140mm 20mm 140mm]{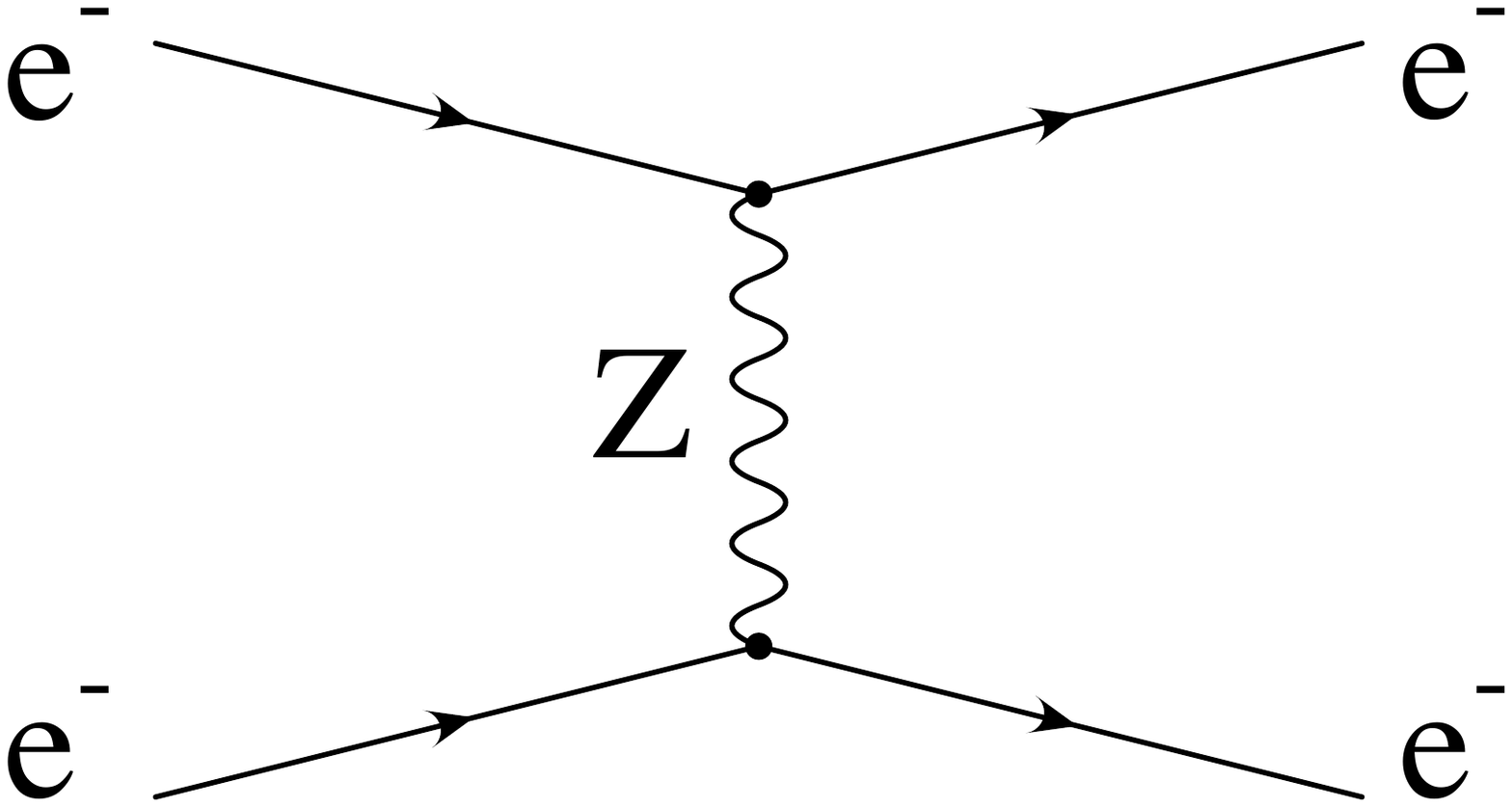}&
\includegraphics[width=3.75cm, trim = 20mm 140mm 20mm 140mm]{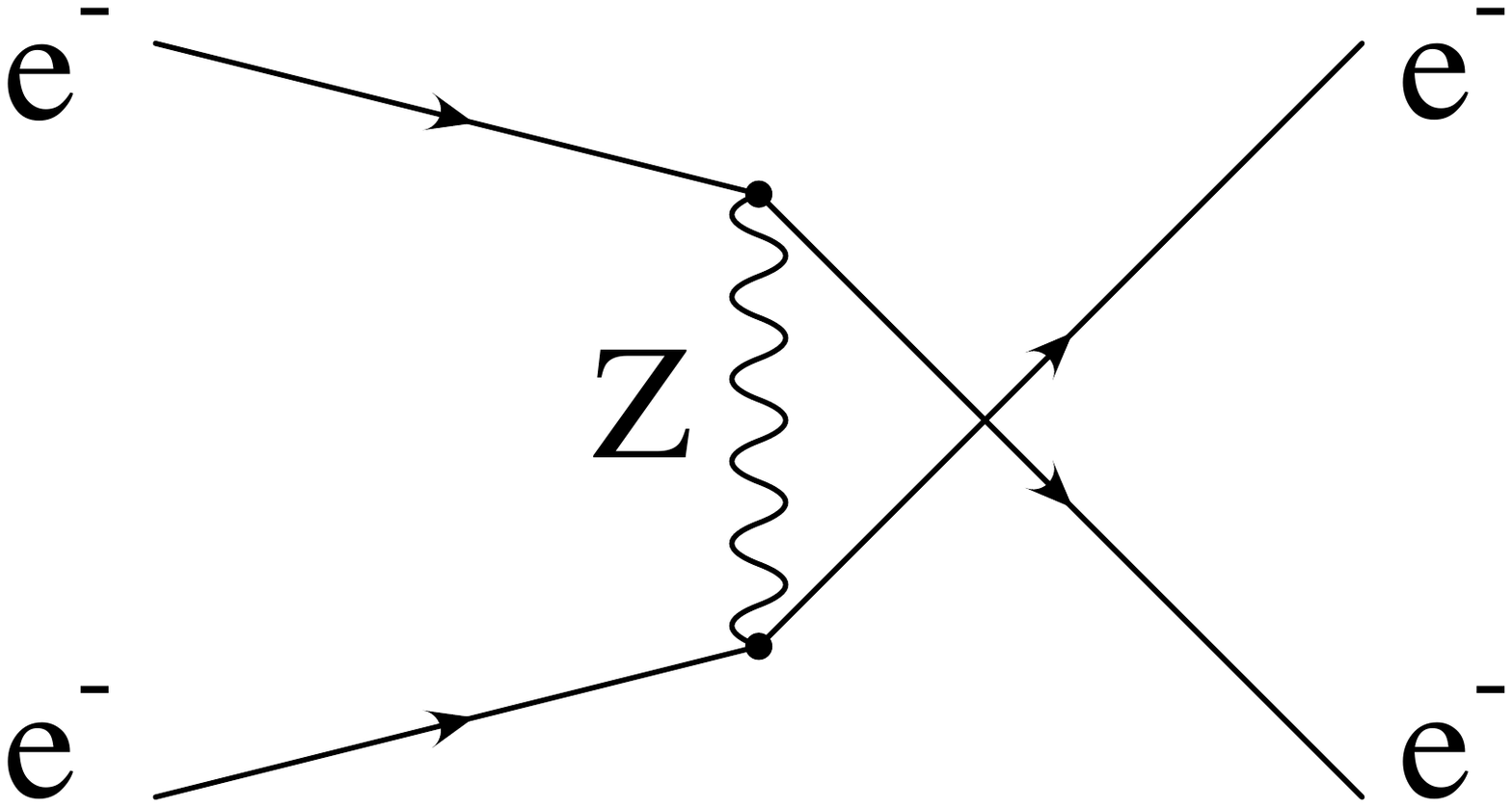}
\end{tabular}
\end{center}
\end{minipage}

\vspace*{15mm}
\noindent 
\caption{{Feynman diagrams for M\o ller scattering at tree level (reproduced from Ref.~\cite{Czarnecki:2000ic})}}
\label{figtree}
\end{figure}

\begin{figure}[htp]
\noindent
\begin{minipage}{16.cm}
\vspace*{5mm}
\hspace*{15mm}
\begin{center}
\begin{tabular}{cccc}
\includegraphics[width=3.75cm, trim = 20mm 140mm 20mm 140mm]{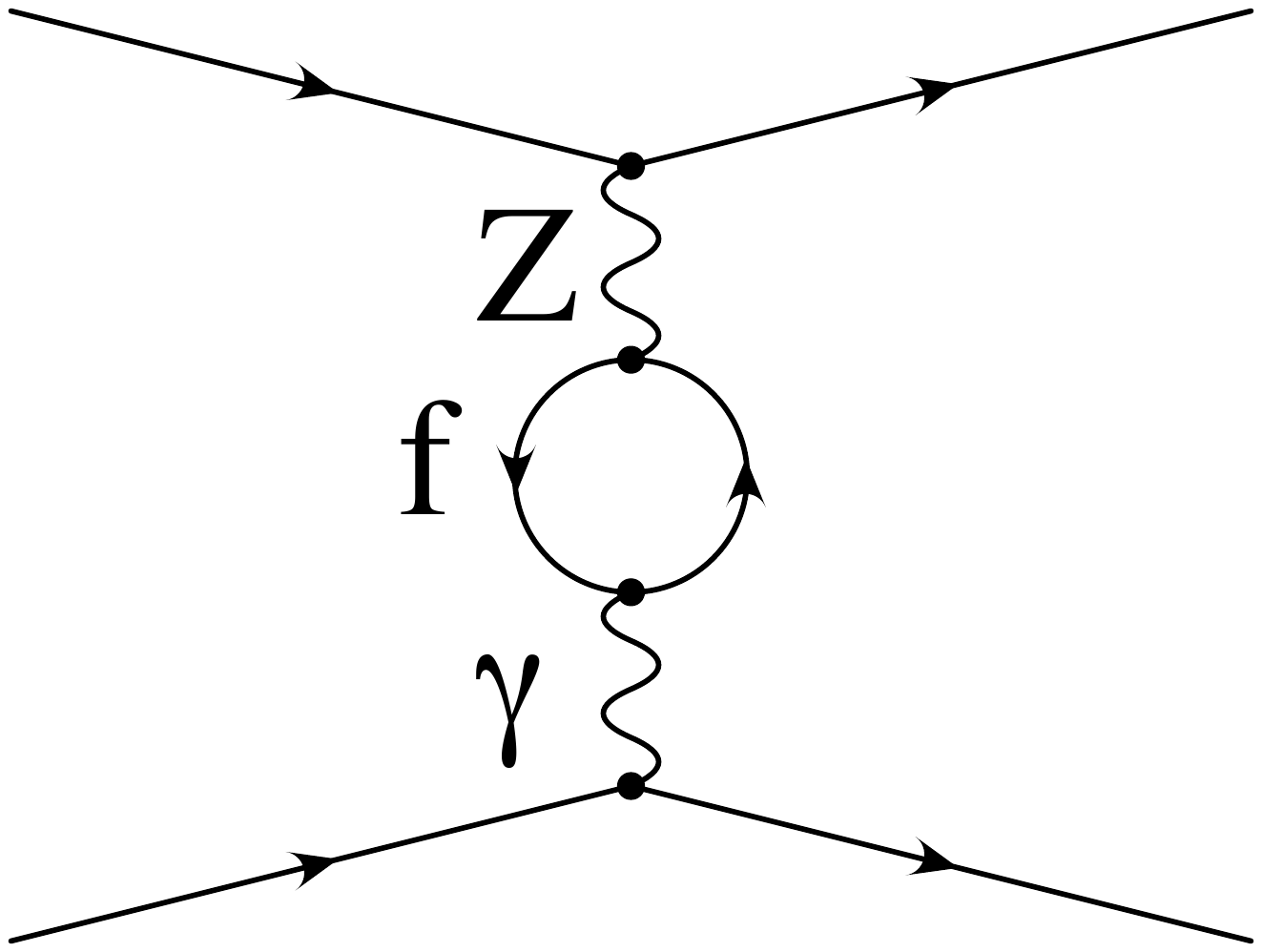}&
\includegraphics[width=3.75cm, trim = 20mm 140mm 20mm 140mm]{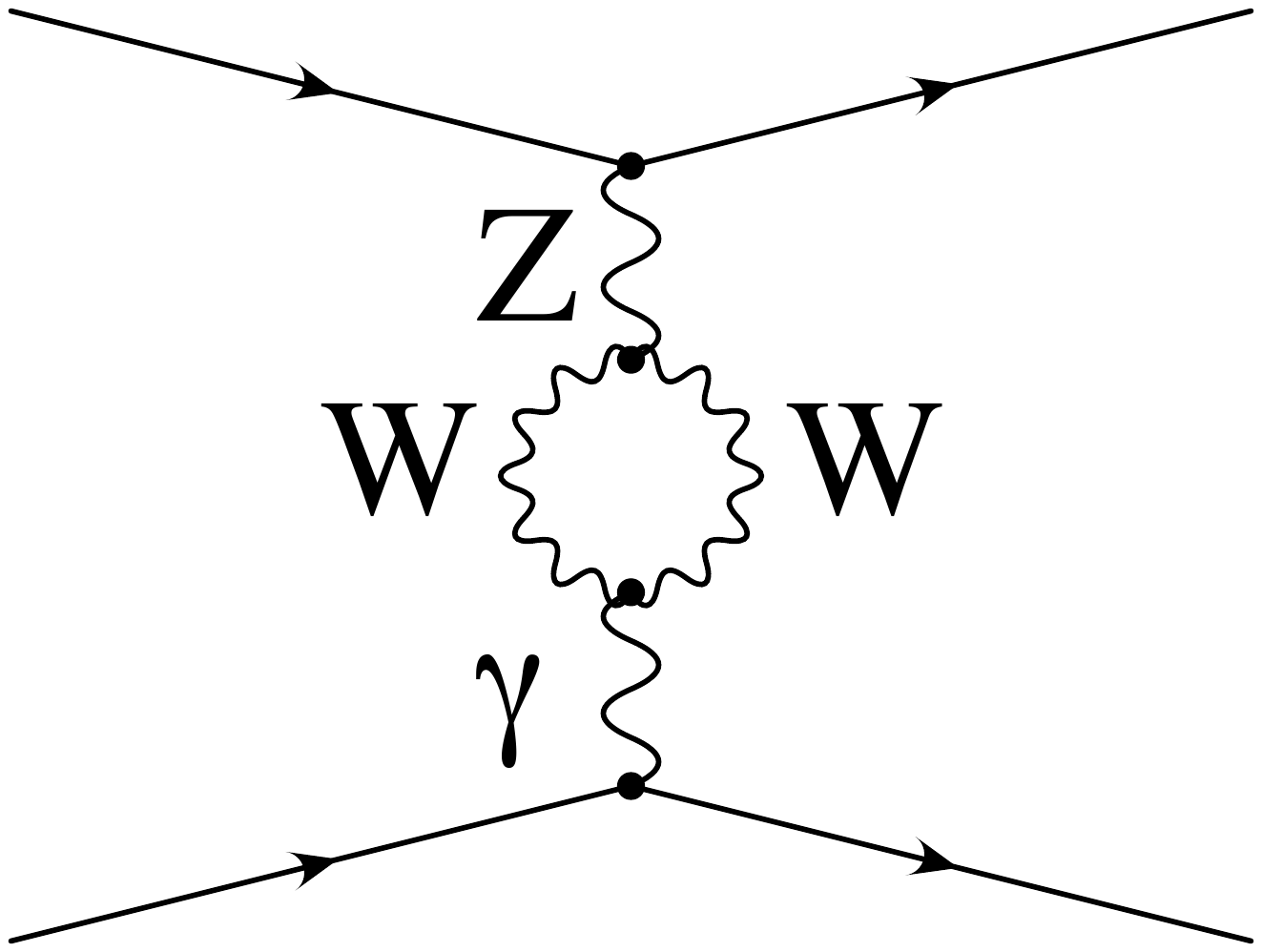}&
\includegraphics[width=3.75cm, trim = 20mm 140mm 20mm 140mm]{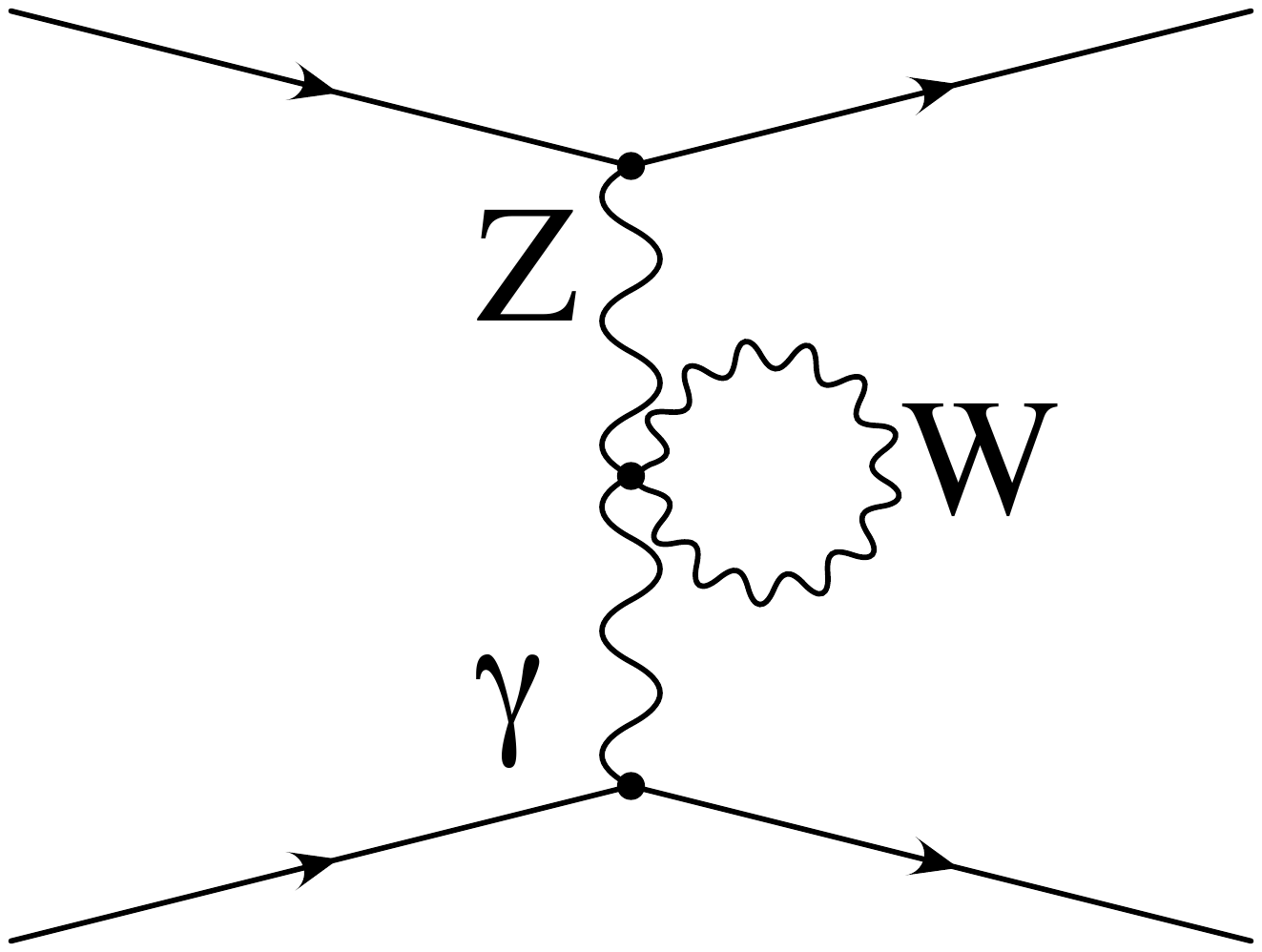}&
\includegraphics[width=3.75cm, trim = 20mm 140mm 20mm 140mm]{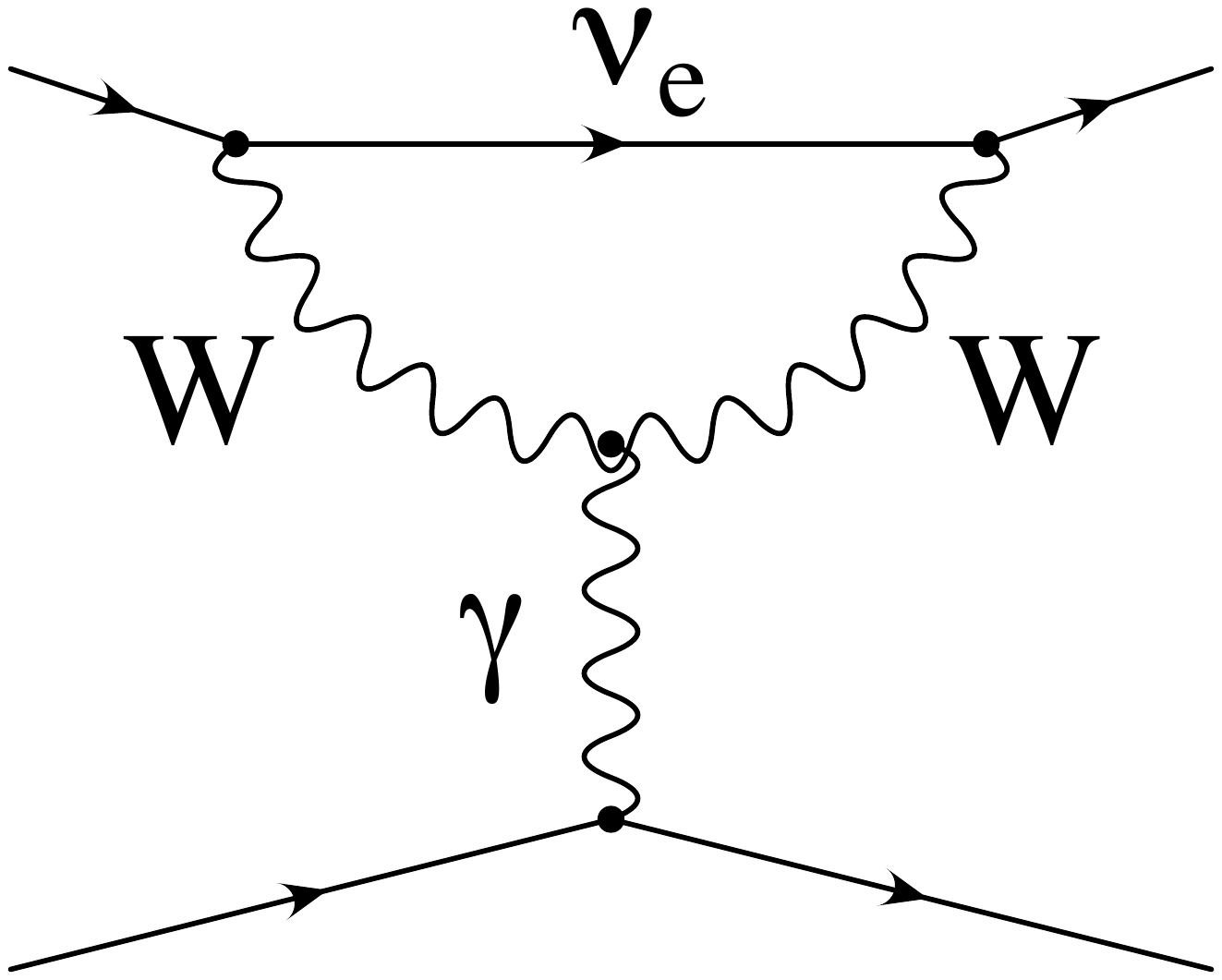}
\end{tabular}
\end{center}
\end{minipage}
\vspace*{15mm}
\noindent 
\caption{{Significant 1-loop radiative corrections: $\gamma-Z$ mixing diagrams and  $W$-loop contribution to
the anapole moment (reproduced from Ref.~\cite{Czarnecki:2000ic})}}
\label{feynmolrad}
\end{figure}

Polarized electron scattering off unpolarized targets provides a clean window to study weak neutral current interactions.  The leading order Feynman diagrams relevant for M\o ller scattering, involving both direct and exchange diagrams that interfere with each other, are shown in Fig.~\ref{figtree}.   The parity-violating asymmetry
in the scattering of longitudinally polarized electrons on unpolarized target electrons $A_{PV}$, due to the interference between the photon and $Z^0$ boson exchange diagrams in Fig.~\ref{figtree}, is given by~\cite{Derman:1979zc}  
\begin{equation}
A_{PV} = {\sigma_R-\sigma_L\over\sigma_R+\sigma_L} = mE{G_F\over\sqrt{2}\pi\alpha}{4\sin^2\theta\over(3+\cos^2\theta)^2}Q^e_W
= mE{G_F\over\sqrt{2}\pi\alpha}\frac{2y(1-y)}{1+y^4+(1-y)^4}Q^e_W
\label {eq:amoll}
\end{equation}
where $Q^e_W$ (proportional to the product of the electron's vector and axial-vector couplings to the $Z^0$ boson) is the weak charge of the electron, $\alpha$ is the fine structure constant, $E$ is the incident beam energy, $m$ is the electron mass, $\theta$ is the scattering angle in the center of mass frame, $y\equiv 1-E^\prime/E$ and $E^\prime$ is the energy of one of the scattered electrons. The electroweak theory prediction at tree level in terms of the weak mixing angle is $Q^e_W = 1 - 4\sin^2\theta_W$; this is modified at the 1-loop level~\cite{Czarnecki:1995fw, Czarnecki:2000ic, Erler:2004in} and becomes dependent on the energy scale at which the measurement is carried out, {\em i.e.} $\sin^2\theta_W$ ``runs". It increases by approximately 3\%\ compared to its value at the scale of the $Z^0$ boson mass, $M_Z$; this and other radiative corrections reduce $Q^e_W$ to $0.0435$, a $\sim 42$\%\ change of its tree level value of $\sim 0.075$ (when evaluated at $M_Z$).
The dominant effect comes 
from the ``$\gamma-Z$ mixing" diagrams depicted in 
Fig.~\ref{feynmolrad}~\cite{Czarnecki:2000ic}.  
The prediction for $A_{PV}$ for the proposed experimental design is 
$\approx 33$~parts per billion (ppb) and 
the
goal is to measure this quantity with an overall precision of 0.7 ppb
and thus achieve a 2.4\%\ measurement of $Q^e_W$. The reduction in the numerical value of $Q^e_W$ 
due to radiative corrections leads to increased fractional accuracy in the determination of the weak mixing
 angle, $\sim 0.1$\%, matching the precision of the single best such determination from measurements of asymmetries in $Z^0$ decays in the $\mathrm{e}^+\mathrm{e}^-$ colliders LEP and SLC. 
An important point to note is that, at the proposed level of measurement accuracy of $A_{PV}$, 
the Standard Model (SM) prediction must be carried out with full treatment of one-loop radiative corrections and leading two-loop corrections.  The current error associated with radiative corrections for MOLLER is estimated to be $\sim 0.2$ ppb, smaller than the expected 0.7 ppb overall precision.  There is an ongoing effort to investigate several classes of diagrams beyond 
one-loop~\cite{n34,n35,n36}, and a plan has been formulated to evaluate the complete set of two-loop corrections at MOLLER kinematics by 2016; such corrections are estimated to be already smaller than the MOLLER statistical error.  The existing work makes it clear that the theoretical uncertainties for the purely leptonic M{\o}ller PV
are well under control, and the planned future work will reinforce that conclusion.

The proposed MOLLER measurement will make a precision (2.4\% relative) measurement of a suppressed Standard Model observable ($Q^e_W \sim 0.0435$) resulting in sensitivity to new neutral current amplitudes as weak as $\sim 10^{-3}\cdot G_F$ from as yet undiscovered dynamics beyond the Standard Model.  The fact that the proposed measurement provides such a sensitive probe of TeV-scale dynamics
beyond the SM (BSM)
is a consequence of a very precise experimental goal ($\sim 10^{-3}\cdot G_F$), the energy scale of the reaction ($Q^2\ll M_Z^2$), and the ability within the 
electroweak theory to provide quantitative predictions with negligible theoretical uncertainty. 
The proposed measurement is likely the only practical way, using a purely leptonic scattering amplitude at 
$Q^2\ll M_Z^2$, to make discoveries in important regions of BSM space 
in the foreseeable future at any existing or planned facility worldwide.

\begin{figure}[!htb]
\begin{center}
\includegraphics[width=6.0in ,angle=0]{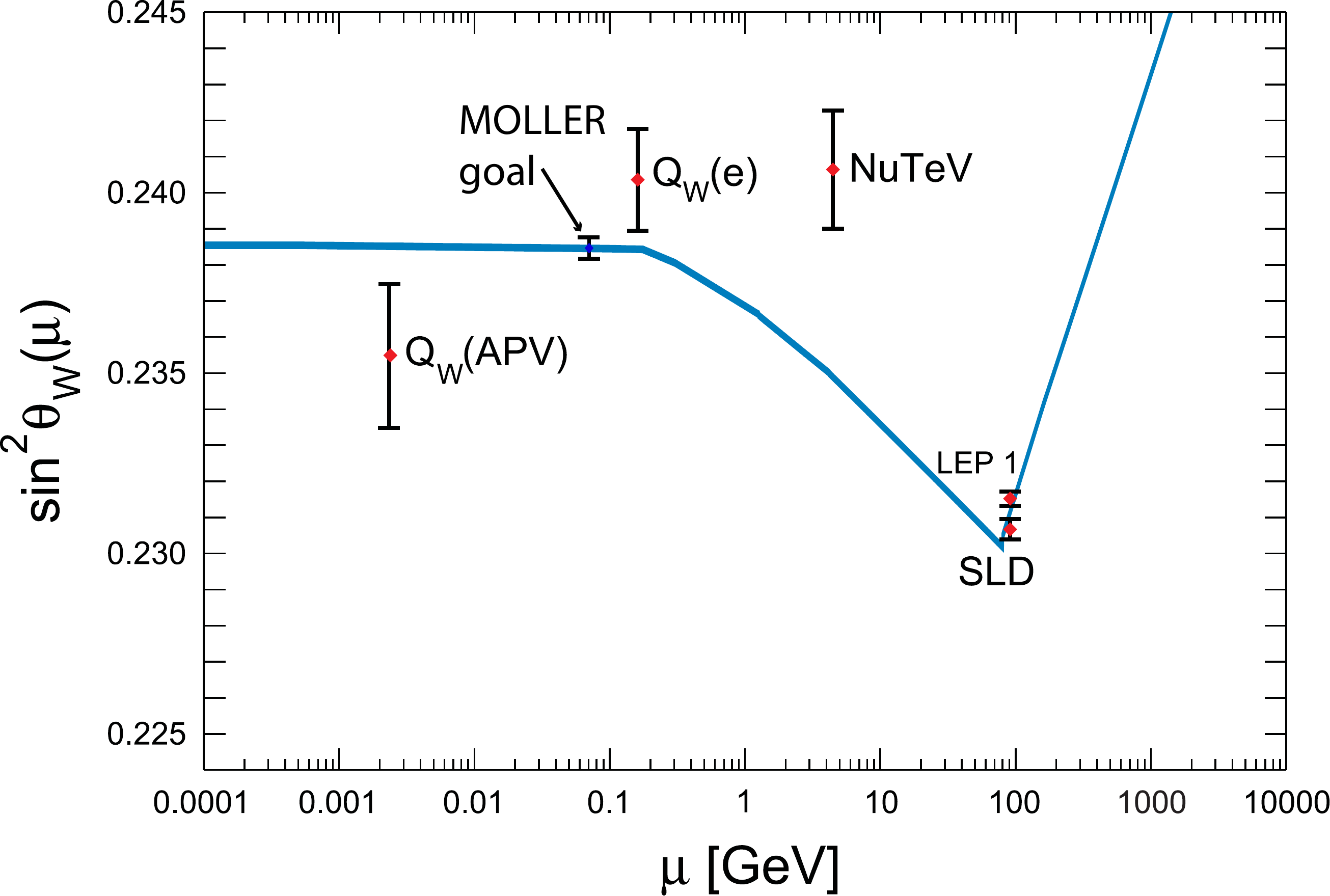}
\end{center}
\caption{The three most precise measurements of the weak mixing angle measurements vs. the energy scale
  $\mu$ are shown as red diamonds with error bars; the curve is reproduced from the PDG~\cite{Amsler:2008zz}. The APV result reflects the reanalysis in Ref.~\cite{csupdate}. The $Q_W(e)$ point is the E158 
result~\cite{Anthony:2005pm}. The NuTeV point is the extracted value from the original publication 
result~\cite{Zeller:2001hh}. The proposed MOLLER measurement is shown at the
appropriate $\mu$ value and the proposed error bar but with the nominal SM prediction as the central value.}
\label{fig:running}
\end{figure}

The {\bf weak mixing angle} $\sin^2\theta_W$ has played a central role in the development and validation of the electroweak theory, especially testing it at the quantum loop level, which has been the central focus of precision electroweak physics over the past couple of decades. To develop the framework, one starts with three fundamental experimental inputs characterizing, respectively,  the strength of electroweak interactions, the scale of the weak interactions, and the level of photon-$Z^0$ boson mixing. 
The three fundamental inputs are chosen to be $\alpha$  (from the Rydberg constant), $G_F$ (from the muon lifetime) and $M_Z$ (from the LEP $Z^0$ line-shape). 
Precise theoretical predictions for other experimental observables at the quantum-loop level can be made if experimental constraints on the strong coupling constant and heavy particle masses, such as $m_H$ and the top quark mass, $m_t$, are also included. 

Precision measurements of the derived parameters such as the W boson mass $M_W$, and the weak mixing angle 
$\sin^2\theta_W$ are then used to test the theory at the level of electroweak radiative corrections. Consistency (or
lack thereof) of various precision measurements can then be used to search for indications of BSM physics. 
One important new development is the discovery of the scalar resonance at LHC with mass of
about 126 GeV. Each individual observable used to extract values for $M_W$ and $\sin^2\theta_W$ can now
be precisely predicted within the SM context.

A crucially important additional feature of MOLLER $A_{PV}$ is that the measurement will be carried out at $Q^2\ll M_Z^2$.
The two best measurements of the weak mixing angle at lower energies are those extracted from the
aforementioned SLAC E158 measurement~\cite{Anthony:2005pm}, 
and the measurement of the weak charge of $^{133}$Cs~\cite{csapv} via studies of table-top atomic parity violation. The interpretation of the latter measurement in terms of an extraction of the weak mixing angle has
been recently updated~\cite{csupdate}. A precise measurement of the weak charge of the proton is expected
from the JLab Qweak experiment via the measurement of $A_{PV}$ in elastic electron proton scattering; the
first result from the commissioning run was recently published~\cite{qweakfirst}.

Since  
$\sin^2\theta_W$ ``runs" as a function of $Q^2$ due to electroweak radiative corrections, one can use $\sin^2\theta_W$ as a bookkeeping parameter to compare the consistency of the full $Q^2$ range
of weak neutral current measurements, as shown in 
Fig.~\ref{fig:running}. The theory error in the low energy extrapolation is comparable to the width of the line in the 
figure~\cite{Erler:2004in}.
MOLLER $A_{PV}$ would be the first low $Q^2$ measurement to match the precision of the single
best high energy measurement at the $Z^0$ resonance. As discussed previously and also further elaborated in the
following with additional examples, low energy measurements have enhanced sensitivity to new physics. 
MOLLER will build on the pioneering low $Q^2$ measurements shown in the figure
to extend the discovery reach for new physics not only to the multi-TeV scale but also, as 
shown in the
following, to light new degrees of freedom. 

\begin{figure}[ht]
\begin{minipage}[b]{0.48\linewidth}
\centering
    \includegraphics[width=3.3in]{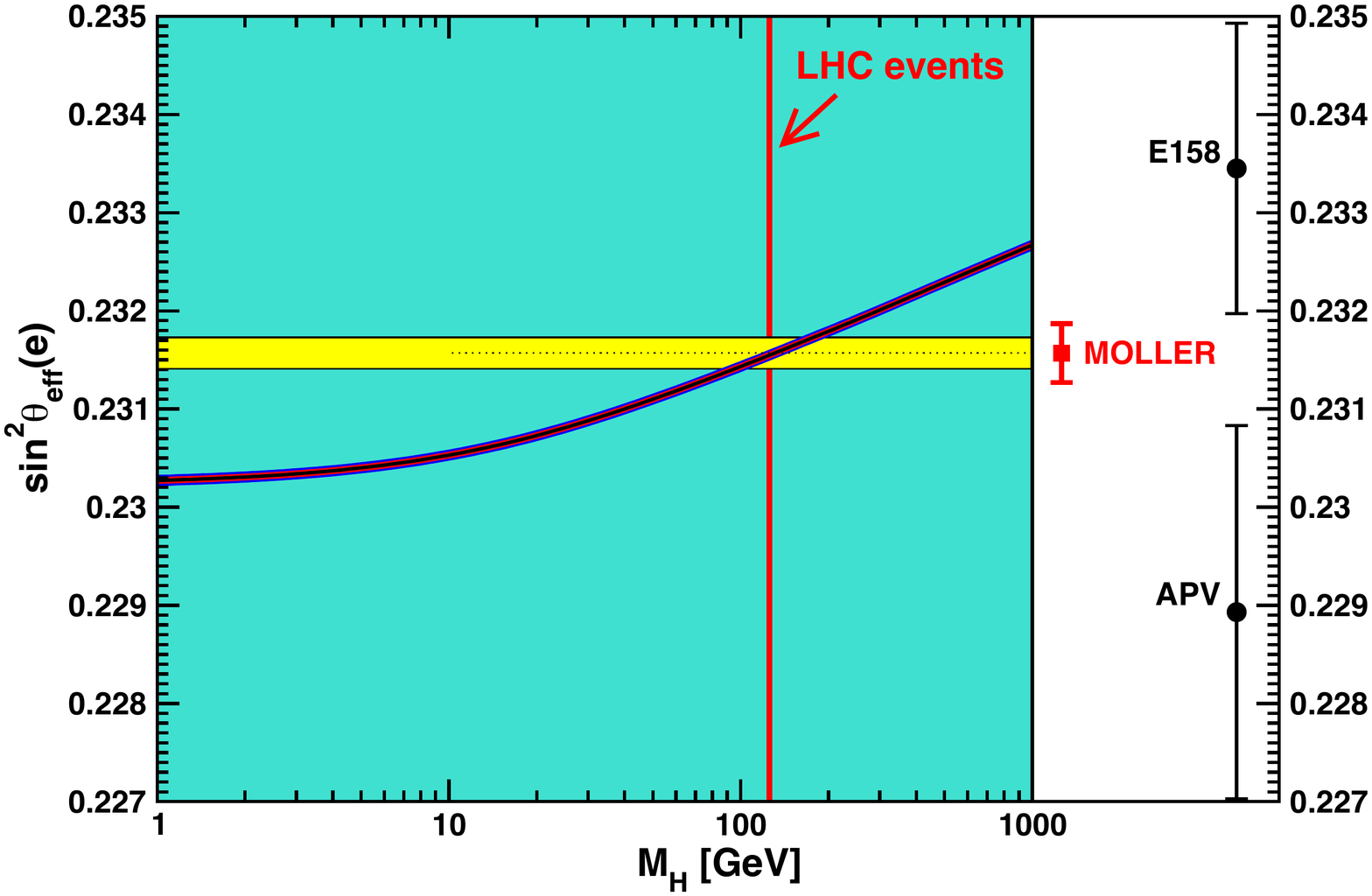}
  \caption{ $\sin^2\theta_W$ vs $m_H$. The yellow band is the world average. 
  The black points are the two most precise measurements
  at $Q^2\ll M_Z^2$. The projected MOLLER error is shown in red. }
  \label{fig:cl:s2twvsmh}
\end{minipage}
\hspace{0.3cm}
\begin{minipage}[b]{0.48\linewidth}
\centering
\includegraphics[width=1.1\linewidth]{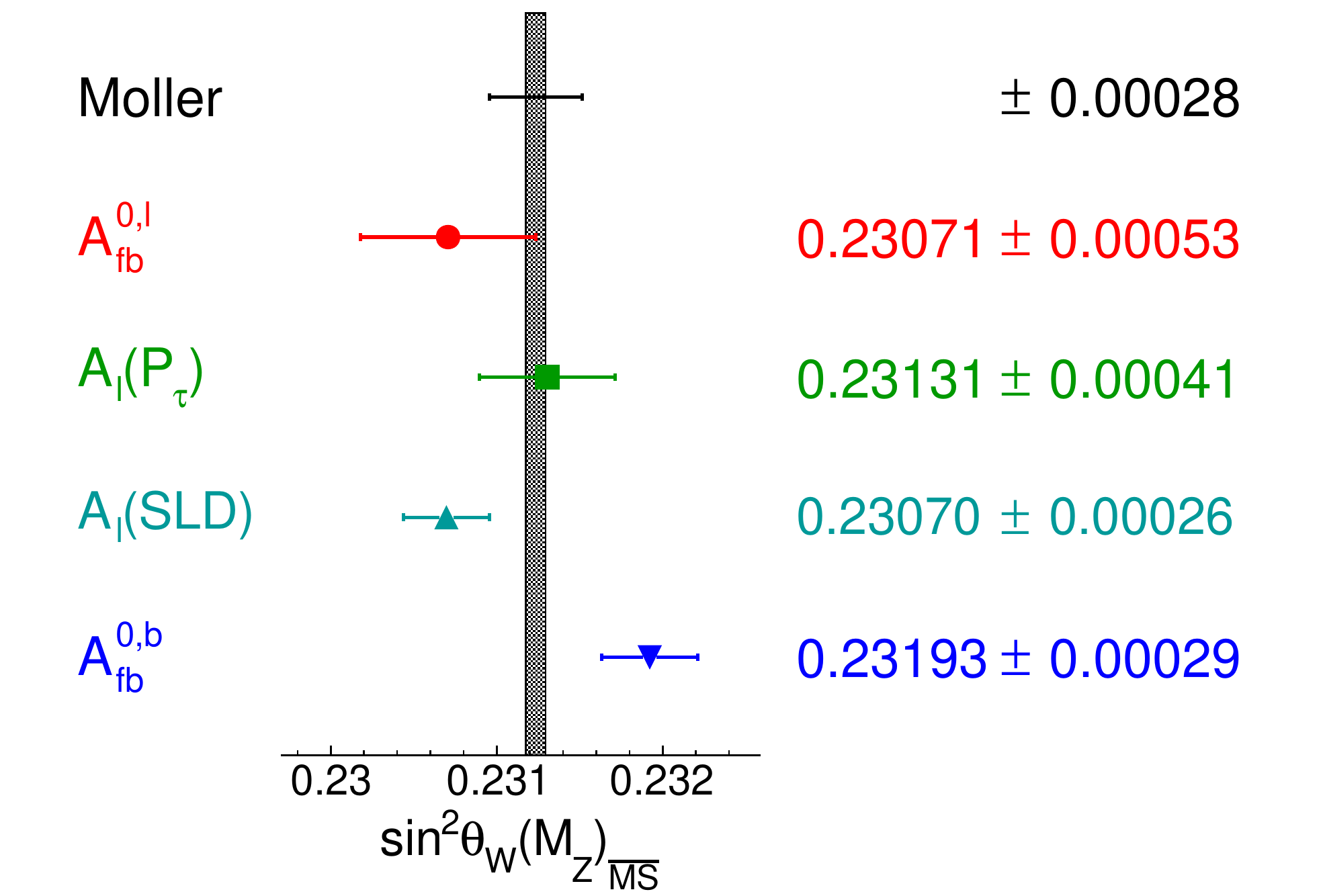}
  \caption{ The four best $\sin^2\theta_W$ measurements and the projected error of the MOLLER proposal.
  The black band represents the theoretical prediction for $m_H = 126$ GeV.}
\label{fig:cl:s2tw}
\end{minipage}
\end{figure}

Figure~\ref{fig:cl:s2twvsmh} shows the dependence 
of $\sin^2\theta_W$ to $m_H$ and the two best published low energy measurements (evolved to $Q\sim M_Z$) discussed above, as
well as the projected $A_{PV}$ error. Remarkably, a variety of BSM dynamics,
such as those discussed in previous sections, can have a significant impact on low $Q^2$ observables while having much reduced impact on corresponding measurements made at colliders. This is because interference effects are highly suppressed on top of the Z$^0$ resonance; those 
measurements account for the bulk of the statistical weight in high energy grand averages.  Since the low energy measurements
at the moment span the entire y-axis range in Fig.~\ref{fig:cl:s2twvsmh}, there
is a  $\pm 10\ \sigma$ discovery potential in the phase space allowed by the existing most precise low energy measurements given the proposed MOLLER $A_{PV}$ uncertainty.

Figure~\ref{fig:cl:s2tw} shows
the four best measurements of $\sin^2\theta_W$ from studies of $Z^{0}$ decays~\cite{:2005ema}
and the projected uncertainty from MOLLER $A_{PV}$. Also shown
is the Standard Model prediction for $m_H = 126$ GeV. 
The bottom two point are the constraints from the most precise single determinations of 
$\sin^2\theta_W$: the left-right asymmetry in $Z$ production at 
SLC ($A_{\mathrm{l}}$(SLD)) and the forward-backward asymmetry in $Z$ decays to b-quarks 
($A_\mathrm{fb}^\mathrm{0,b}$). Each of the two measurements taken independently implies very 
different BSM dynamics~\cite{Marciano:2006zu}.

The proposed MOLLER $A_{PV}$ measurement would achieve a sensitivity of 
$\delta(\sin^2\theta_W) = \pm 0.00028$.  That is the most precise anticipated weak mixing angle measurement currently proposed over the next decade at low or high energy.  The most precise proposed weak charge measurement is the Mainz MESA P2 proton weak charge measurement with anticipated precision $\delta(\sin^2\theta_W) = \pm 
0.00034$.
Proposals for anti-neutrino scattering, both deep-inelastic~\cite{nusong:2009} and elastic~\cite{Conrad:2005}, also fall short of the MOLLER projection.  In particular, elastic anti-neutrino-electron scattering is the best direct comparison to MOLLER as a purely leptonic low $Q^2$ measurement.  
Reactor
experiment projections have fallen 
short of the proposed MOLLER goal.  Matching MOLLER precision and accuracy likely would require beta-beams or neutrino factories.  Finally, the 
projected uncertainty from forward-backward asymmetries after 300 fb$^{-1}$ integrated luminosity at the LHC is a systematics limited $\delta(\sin^2\theta_W) = \pm 0.00036$, with the dominant error being from parton distribution function (pdf) uncertainties~\cite{snomass:2013}.

A fairly general and model-independent way to quantify the energy scale of BSM high-energy dynamics (that MOLLER is sensitive to) is to express the resulting new amplitudes at low energies in terms of {\bf contact interactions} (dimension-6
non-renormalizable operators) among leptons and quarks~\cite{Eichten:1983hw}.  
Specializing here to vector and axial-vector interactions between electrons and/or positrons, 
the interaction Lagrangian is
characterized by a mass scale $\Lambda$ and coupling constants $g_{ij}$ labeled by the chirality of the leptons.  
For the MOLLER $A_{PV}$ measurement with 2.4\% total uncertainty (and no additional theoretical uncertainty) the resulting sensitivity  to new 4-electron contact interaction amplitudes can be expressed as:
\begin{equation}
{\Lambda\over \sqrt{|g_{RR}^2 - g_{LL}^2|}} = {1\over \sqrt{\sqrt{2} G_F |\Delta Q_W^e|}} \simeq
{246.22 \mbox{ GeV}\over \sqrt{0.023 Q_W^e}} = 7.5 \mbox{ TeV}.
\label{reach}
\end{equation}
For example, models of lepton compositeness are characterized by strong coupling dynamics. Taking $\sqrt{|g_{RR}^2 - g_{LL}^2|} = 2 \pi$ shows that mass scales as large as $\Lambda = 47$~TeV can be probed, far beyond the center of mass energies of any current or planned high energy accelerator. This allows electron substructure to be studied down to the level of $4\times 10^{-21}$~m.

The MOLLER measurement is sensitive to Beyond the Standard Model (BSM) scenarios that predict observable consequences at the LHC and those that might escape detection there.  
If the {\bf LHC observes an anomaly} in the next decade, then MOLLER will have the sensitivity to be part of a few select measurements that will provide important constraints to choose among possible BSM scenarios to explain the anomaly.  One example of such a scenario are new, {\bf super-massive $Z^\prime$ bosons} with masses in the multi-TeV range, as predicted in many BSM theories.  The MOLLER $A_{PV}$ measurement would see a a statistically significant deviation in many models that predict $Z^\prime$ bosons in the 1--3~TeV mass range.  Specific examples have been considered in ~\cite{newjenszprime} for a fairly large class of family-universal models contained in the $E_6$ gauge group.  Should a $Z^\prime$ resonance in the 1--3 TeV range be found at the LHC,  the importance of off-peak LHC data as well as low-energy precision EW data to completely disentangle all  of the chiral $Z^\prime$ couplings to SM particles has been emphasized~\cite{petriello}, with MOLLER providing important constraints.  Another important class of BSM physics that could have signatures both at the LHC and in MOLLER are new particles predicted by the {\bf Minimal Supersymmetric Standard Model} observed through radiative loop effects (R-parity conserving) or tree-level interactions (R-parity violating)~\cite{Kurylov:2003zh,RamseyMusolf:2006vr}.  The RPV and RPC models generate effects of opposite sign in the weak charge.   The
 difference is not academic, since RPC would imply that the lightest supersymmetric particle is stable and therefore
 an obvious candidate for the non-baryonic dark matter which is needed to understand 
galactic-scale dynamics. On the other hand, RPV would imply that neutrinos are Majorana particles. 

\begin{figure}[htb]
  \begin{center}
 \includegraphics[width=6.in]{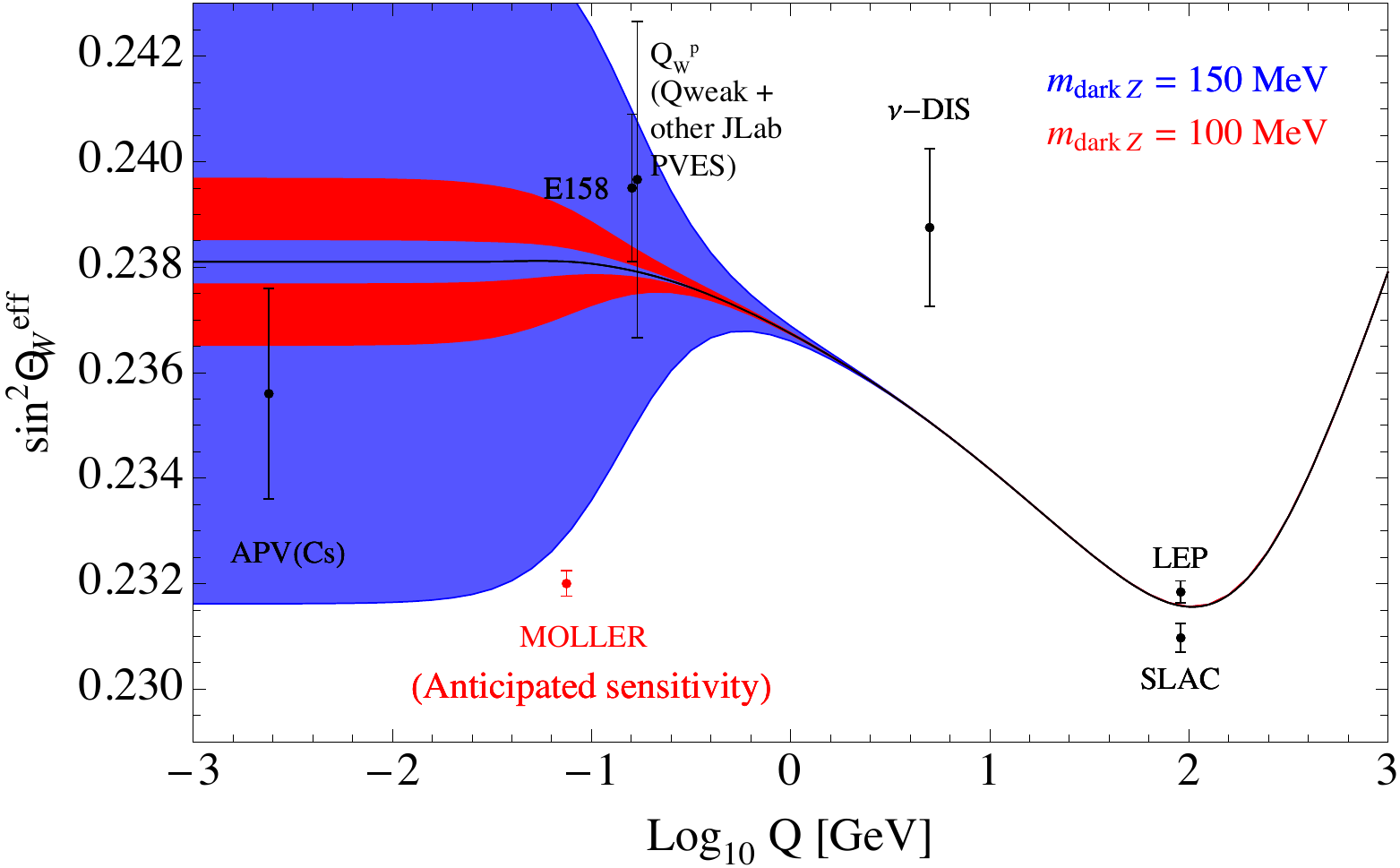}
  \end{center}
  \caption{The potential deviations of $\sin^2\theta_W(Q)$ under the scenario of small 
  parity-violating admixture of a light $Z_d$ to explain the $(g-2)_\mu$ anomaly is shown for two different masses, taking into account rare kaon decay constraints. Also shown is the extracted
  $\sin^2\theta_W$ from the value of the weak charge of the proton $Q_W^p$ quoted in the recently published  Qweak result from the commissioning run~\cite{qweakfirst}.  The proposed MOLLER measurement is shown at the
  appropriate $Q$ value and the proposed error bar but with an arbitrary central value. Note: the sign of the
  deviation is model-dependent.}
\label{fig:darkz}
\end{figure}

 If the {\bf LHC continues to agree with the Standard Model} wtih high luminosity running at the full 14 TeV energy, then MOLLER will be a significant component of a global strategy to discover signatures of a variety of physics that could escape LHC detection.  One example is hidden weak scale scenarios such as {\bf compressed supersymmetry}~\cite{Chao:2014}.  Such a situation would occur if one of the superpartner 
masses is relatively light, as would be the case if the super-partner masses were nearly degenerate. In that 
scenario, the LHC signatures would be very challenging to disentangle 
from QCD backgrounds.  Another example is lepton number violating amplitudes mediated by  {\bf doubly charged scalars}.  
The MOLLER measurement is one of the rare low $Q^2$ observables with sensitivity to such amplitudes, which naturally arise in extended Higgs sector models containing complex triplet 
representations of SU(2).  In a left-right symmetric model, for example, the proposed MOLLER measurement would lead to the
most stringent probe of the left-handed charged scalar and
its coupling to electrons, with a reach of
$${M_{\delta_L}\over |h_L^{ee}|} \sim 5.3\mbox{ TeV},$$
significantly above the LEP 2 constraint of about 3~TeV. Moreover, such sensitivity is complementary to other 
sensitive probes such as lepton-flavor violation and neutrinoless double-beta decay searches~\cite{Cirigliano:2004mv}.
Finally,  the interesting possiblity of a light MeV-scale dark matter mediator known as the {\bf ``dark'' Z}~\cite{Davoudiasl:2012ag,Davoudiasl:2012qa} has been recently investigated.  It is denoted as 
$Z_d$ and of mass $m_{Z_d}$,  and it stems from a spontaneously broken $U(1)_d$ gauge symmetry associated with
a secluded ``dark" particle sector. The $Z_d$ boson can couple to SM particles through
a combination of kinetic and mass mixing with the photon and 
the $Z^0$-boson, with couplings $\varepsilon$ and $\varepsilon_Z = \frac{m_{Z_d}}{m_Z}\delta$ respectively. 
In the presence of mass mixing ($\delta \neq 0$),  a new source of ``dark'' parity violation arises~\cite{Davoudiasl:2012ag} such that it has negligible effect on other precision electroweak observables at high energy, 
but is quite discernable at low $Q^2$ through a shift in the weak mixing angle~\cite{Davoudiasl:2012qa}.
Recently, it has been pointed out~\cite{Davoudiasl:2014kua} that the existing constraints 
are considerably weakened if the ``dark" $Z$ decays to other dark matter particles, rendering
the branching ratio $Z_d\rightarrow e^+e^-\ll 1$. In such a scenario, the only constraints
on $Z_d$ masses in the range between 50 and 200 MeV would come from neutral current
parity-violation measurements 
and rare kaon decay experiments 
($K\rightarrow\pi + Z_d$, $Z_d\rightarrow$ missing energy). Figure~\ref{fig:darkz} 
shows the range of possible deviations to
$\sin^2\theta_W(Q)$ for $Z_d$ mass of 100 and 150 MeV, under the scenario that the 
``dark" Z explains the $(g-2)_\mu$ anomaly, but taking into account constraints
from the K decay measurements. It can be seen that the proposed MOLLER $A_{PV}$
measurement has significant discovery potential under this scenario.

In summary, the discovery reach of the proposed MOLLER measurement 
is unmatched  by any proposed experiment measuring a flavor- and CP-conserving process  over the next decade.
It results in a unique window to new physics at MeV and multi-TeV scales, complementary to direct searches at high energy colliders such as the Large Hadron Collider (LHC).

\section{Experimental Overview}

\begin{figure}[!htb]
  \begin{center}
\includegraphics[width=6.0in]{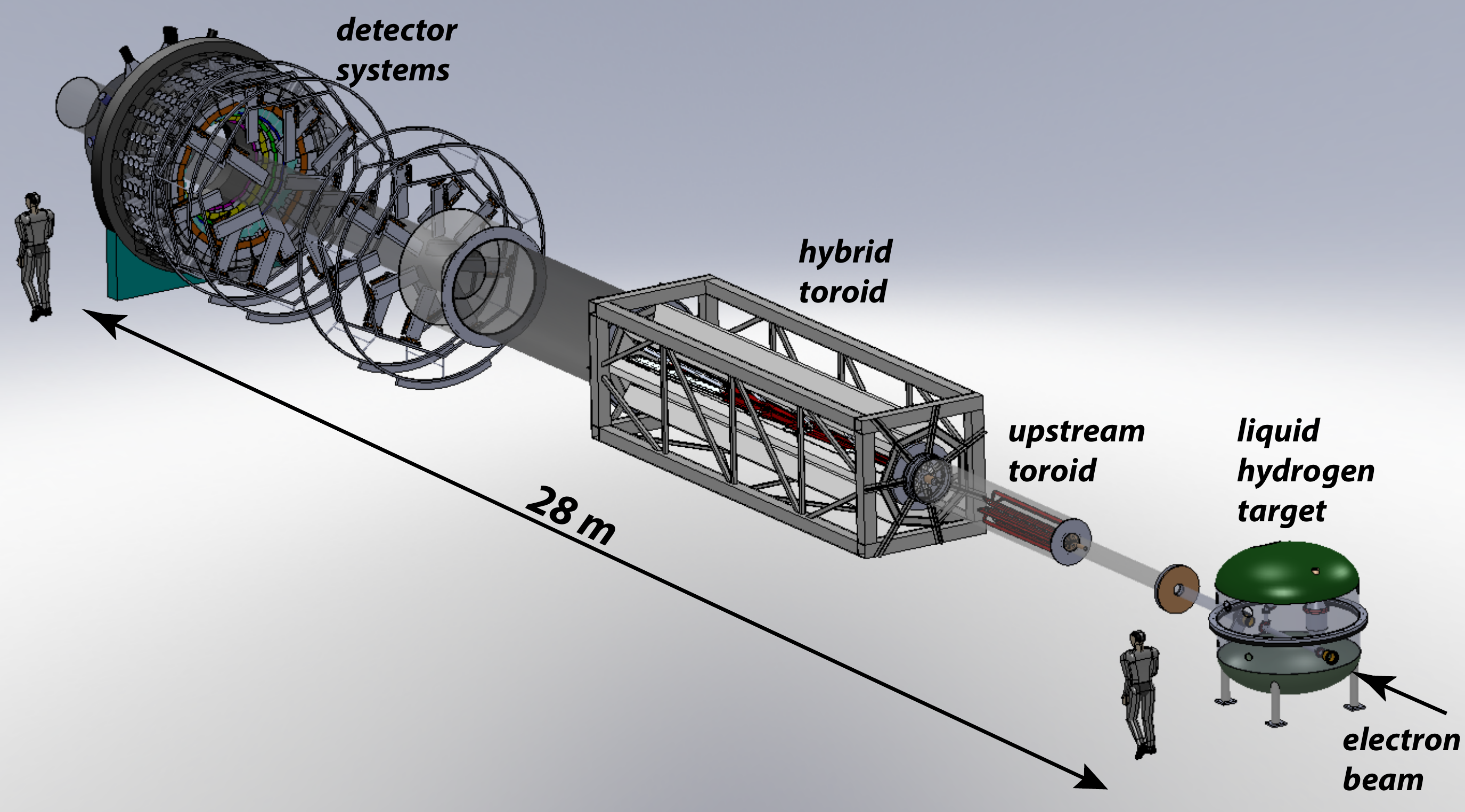}
  \end{center}
  \caption{MOLLER Experiment Overview: Layout of the target, spectrometer and detectors.}
  \label{fig:halla}
\end{figure}

In this section, 
a brief overview of the MOLLER experimental design is given.  
The experimental design is driven by the need to measure a very small parity-violating asymmetry which requires measurement of  the scattered electron flux at an unprecedently high rate and careful attention to a range of systematic effects.  The MOLLER design is grounded on the extensive experience gained by the collaboration from other high flux
integrating (as opposed to counting individual particles) parity-violation measurements such as MIT-Bates $^{12}$C~\cite{Souder:1990ia},
SAMPLE~\cite{Spayde:2003nr}, HAPPEX~\cite{Acha:2006my}, SLAC E158~\cite{Anthony:2005pm}, PREX~\cite{PREX}, and Qweak~\cite{Qweak}.
A CAD-generated rendition of the layout of the MOLLER apparatus to be placed in Hall A at JLab is shown
in Fig.~\ref{fig:halla}, where a 11 GeV longitudinally polarized electron beam would be incident on the 1.5 m liquid hydrogen target. M\o ller electrons (beam electrons scattering off target electrons) in the full range of the azimuth and spanning the polar angular range 5 mrad $<\theta_{lab}<$ 17 mrad, would be separated from background and brought to a ring focus $\sim 30$~m downstream of the target by a spectrometer system consisting of a pair of toroidal magnet assemblies and precision collimators. 
The M\o ller ring would be intercepted by a system of quartz detectors; the resulting Cherenkov light would provide a relative measure of the scattered flux. The experimental techniques for producing an ultra-stable polarized electron beam, systematic
control at the part per billion level, calibration techniques to control normalization errors including the degree of electron 
beam polarization at the 1\%\ level have been continuously improved over fifteen years of development at JLab.

The {\bf polarized electron beam} required is 75 $\mu$A with 80\% longitudinal polarization (or 60 $\mu$A with 90\% longitudinal polarization would achieve the same figure of merit and be more compatible with running of the other JLab experimental halls).  This is easily achievable based on the recent experience with the Qweak experiment where $\sim 180\ \mu$A of $\sim 89\%$ polarized beam was routinely delivered by the polarized source.  The electron beam's helicity is planned to be reversed at a rapid (1.92 kHz) rate to cancel out the effect of slow drifts.  Corrections must be made for any correlation of the beam properties (position, angle, energy) with respect to the electron beam helicity.  Procedures for doing this have been well developed in previous experiments.  Averaged over the entire data collection period, the beam trajectory must remain unchanged with
respect to the sign of the electron beam polarization at the sub-nanometer level in order to keep beam-related false asymmetry corrections at the 1 ppb level.  The goals for these corrections and their errors should be achieveable with modest upgrades to the beamline instrumentation and continued development of the polarized source and accelerator setup and control procedures that worked well in previous experiments.
It will be necessary to use a ``slow reversal'' of beam helicity (such as an optically inserted half-wave plate, spin flip in the polarized injector with a ``double-Wien'' system, and ``$g-2$'' spin flip with an accelerator energy change) to further
cancel systematic errors to the 0.1 ppb level, from sources such as residual electronics cross-talk and  higher-order effects such as potential helicity-dependent
variations in the beam spot size.

The experiment requires {\bf precision electron beam polarimetry} at the level of 0.4\%.  In order to reach a robust 0.4\% precision, 
a Compton polarimeter 
will be used
for a continuous measure of beam polarization.  Independent analysis of scattered photons and electrons provides a pair of continuous measurements with a high degree of independence in systematic errors. This polarimeter will be cross checked against periodic measurements with a M{\o}ller polarimeter using ferromagnetic foil targets. An upgrade to the Hall A  M{\o}ller polarimeter, presently underway, will support improvements and studies that should ultimately lead to a systematic accuracy near 0.4\%.   An alternative second-stage upgrade would incorporate a polarized atomic hydrogen gas target in the M{\o}ller polarimeter, which would provide for continuous operation of a polarimeter with systematic uncertainties completely independent of the Compton-scattered photon or electron measurements. 

In order to achieve the necessary rate, the {\bf liquid hydrogen target} is planned to be 150 cm long.  This requires a cryogenic target system capable of handling a heat load of $\sim 5$~kW from the beam. This would be the highest power liquid hydrogen target constructed, but it would be based on successful experience with the operation of the Qweak target which successfully operated up to 180 $\mu$A with a total power of 2.9 kW~\cite{Smith:2012}.  The final design of the MOLLER target will make use of computational fluid dynamics (CFD), a key recent
development which has been validated by the successful operation of the Qweak target.  From the physics point of view, the most
important design consideration is suppression of density fluctuations at the timescale of the helicity flip rate,
which can ruin the statistical reach of the
flux integration technique. Preliminary estimates based on operational experience with the Qweak target~\cite{Smith:2012}
suggest that density variation can be maintained
 at $\lesssim 26$ ppm at 1.92 kHz (compared to the expected counting statistics width of $\sim 83$ ppm/pair at 75 $\mu$A), corresponding to accepatable 5\%\ excess noise.

\begin{figure}[!htb]
  \begin{center}
  \includegraphics[width=5.0in]{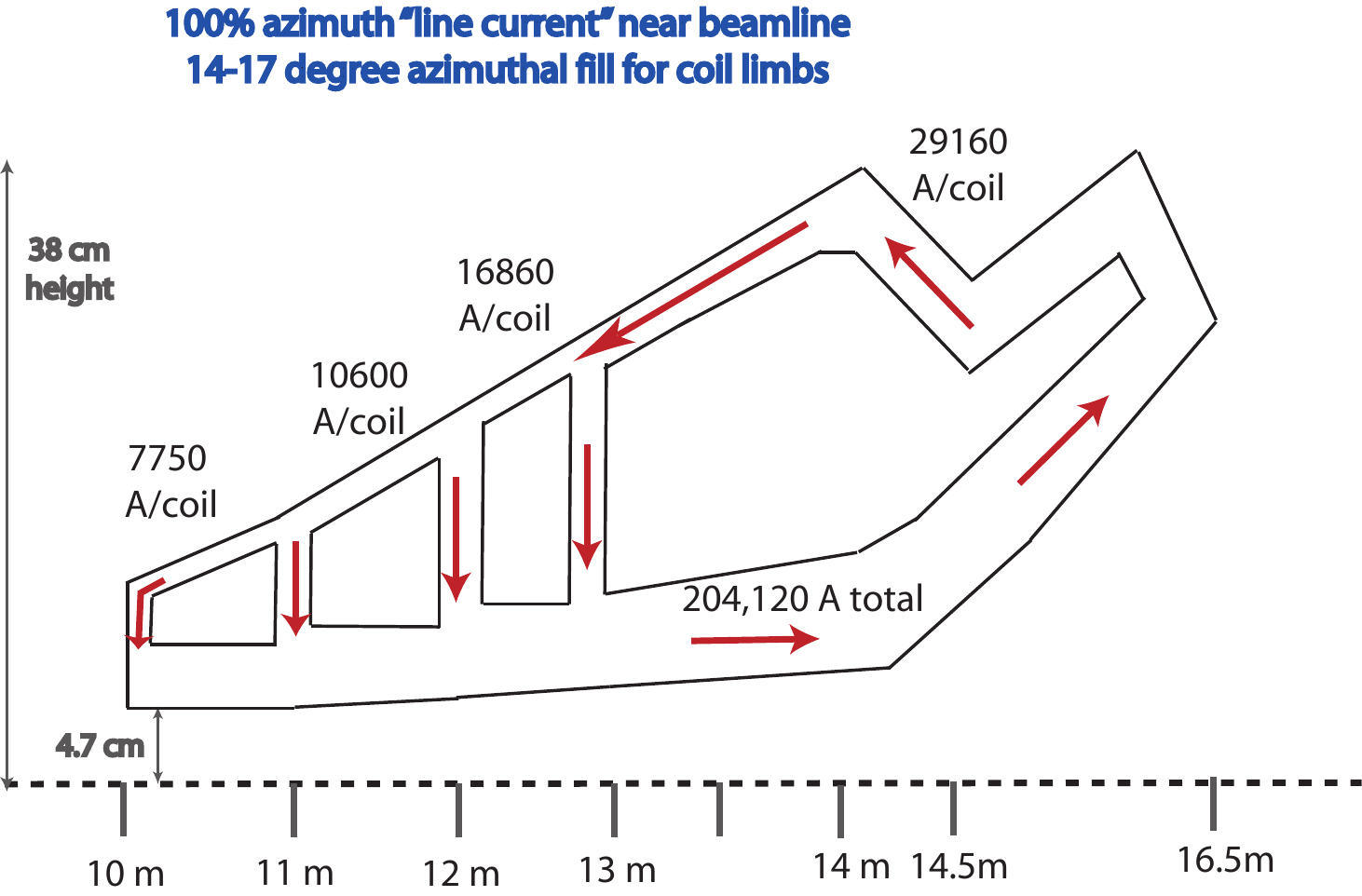}
  \end{center}
  \caption{Schematic of the hybrid toroid design concept.}
  \label{fig:hybrid}
\end{figure}

\begin{figure}[!htb]
  \begin{center}
  \includegraphics[width=5.0in]{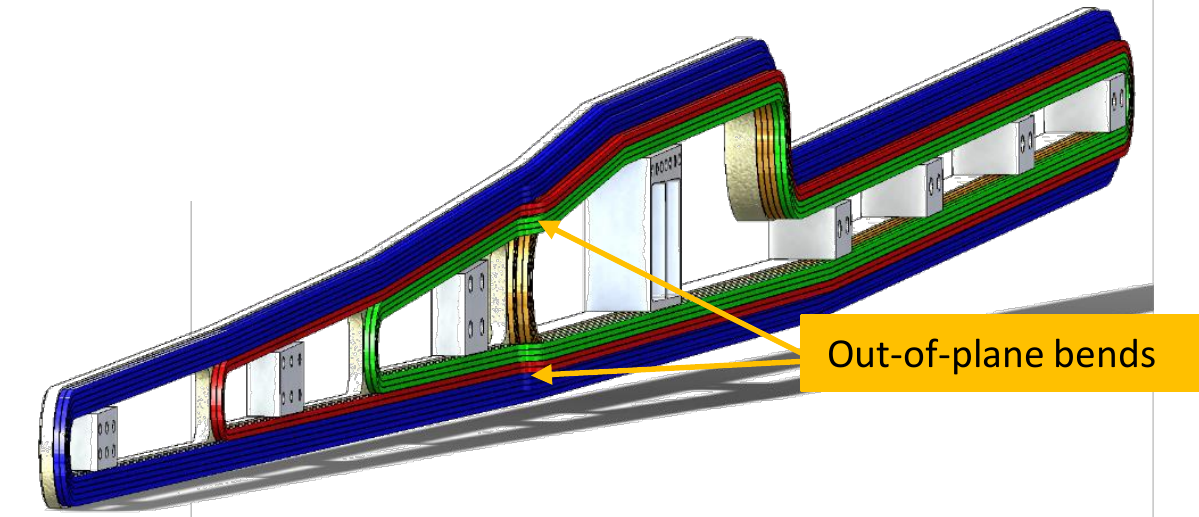}
  \end{center}
  \caption{Preliminary engineering design of a single hybrid coil with part of the encapsulant removed to show the return paths and out of plane bends.}
  \label{fig:hybrid_engineered}
\end{figure}

A {\bf precision collimation system} carefully designed to minimize backgrounds will accept all M{\o}ller scattered electrons in the polar angle range $\Theta_{COM} = 60^\circ - 120^\circ$ (corresponding to a lab polar scattering angle range of 5 mrad $<\theta_{lab}<$ 17 mrad).  The {\bf spectrometer system} that focusses these scattered particles is designed to achieve two goals: 100\% azimuthal acceptance and the ability to focus the scattered M{\o}ller flux over a large fractional momentum bite with adequate separation from backgrounds.  These considerations have led 
to a unique solution involving two back-to-back sets of toroidal coils, one of them of conventional geometry (albeit long and quite skinny) while the other is of quite novel geometry.  Due to the special nature of identical particle scattering, it is possible to achieve 100\% azimuthal acceptance in such a system by choosing an odd number of coils.  The idea is to accept both forward
and backward (in center of mass angle) M\o llers in each $\phi$ bite.  Since these
are identical particles, those that are accepted in one $\phi$ bite also represent all the statistics available
in the $\phi$ bite that is diametrically opposed ($180^\circ + \phi$), which is the sector that is blocked due to the presence of a toroidal coil.  An event with a forward angle scattered M{\o}ller electron that azimuthally scatters into a blocked sector is detected via its backward angle scattered partner in the open sector diametrically opposed, and vice versa.  The focussing and separation of the scattered M{\o}ller electrons is challenging due to their large scattered energy range $E^\prime_{lab} = 1.7 - 8.5$ GeV and the need to separate them from the primary background of elastic and inelastic electron-proton scattering.  The solution is a combination of two toroidal magnets which together act  in a non-linear way on the charged particle trajectories.
The first is a conventional toroid placed 6 m downstream of the target and the second, a novel ``hybrid'' toroid placed between 10 and 16 m downstream of the target.  Each of the two toroidal fields is constructed out of seven identical coils uniformly spaced in the azimuth.  The ``hybrid'' toroid has several novel features to provide the required field to focus the large range of electron scattering angles and momenta.  It has four current return paths, as shown in Fig.~\ref{fig:hybrid} and some novel bends that minimize the field in certain critical regions.  A preliminary engineering design of this hybrid toroid with realistic conductor, water cooling circuits, coil carriers and support frame has been produced.  The design for a single coil is shown in Fig.~\ref{fig:hybrid_engineered}.

\begin{figure}[!htb]
\begin{center}
\includegraphics[width=1.0\linewidth]{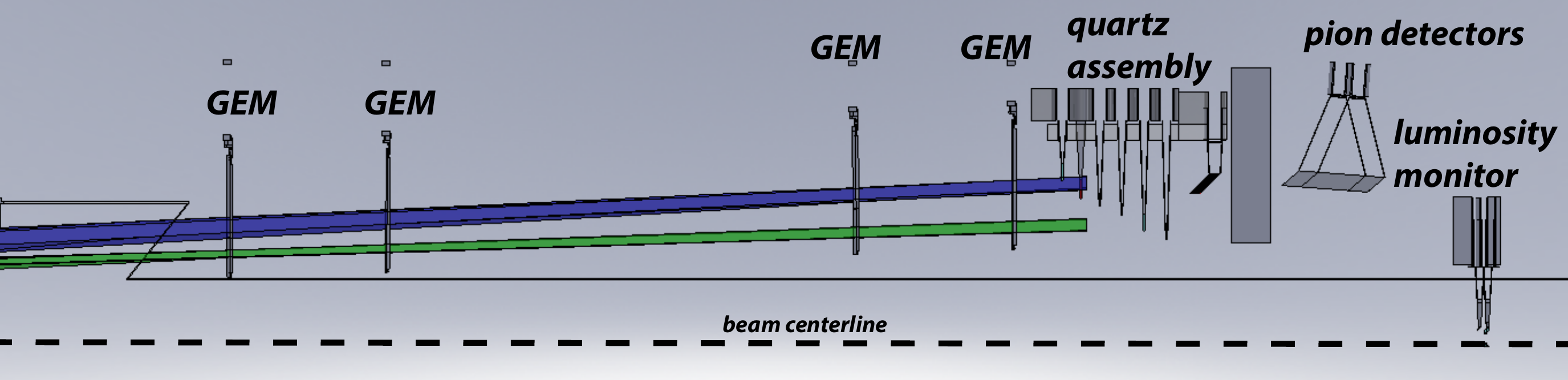}
\end{center}
\caption{Layout of the main integrating and tracking detectors. Predicted trajectories from elastically scattered electrons from target protons (green) and target electrons (blue) are also shown.}
\label{detectoroverview}
\end{figure}

The MOLLER apparatus consists of a number of {\bf detector systems}: 
integrating (current mode) detectors, for the
asymmetry measurements of both signal and background, and beam and target monitoring, as well as tracking (counting mode) detectors for
spectrometer calibration, electron momentum distribution and background measurements.
An overview of the main detector systems is shown in
Fig.~\ref{detectoroverview}. The toroidal spectrometer will focus the M\o ller electrons
$\approx$ 28 m downstream of the target center onto a ring with a central radius of $\approx$ 100 cm and a radial spread of $\approx$ 10 cm.  The region between a radius of 60 to 110 cm will be populated
by a series of detectors with radial and azimuthal segmentation. These detectors will measure $A_{PV}$ for M{\o}ller
scattering and, equally important, will also measure $A_{PV}$ for the irreducible background processes of elastic and
inelastic electron proton scattering. Detectors at very forward angle will monitor window to window fluctuations in the scattered flux
for diagnostic purposes. Lead-glass detectors placed behind the main M{\o}ller ring detectors and shielding, combined
with two planes of gas electron multipliers (GEMs) will measure hadronic background dilutions and asymmetries.
Finally, four planes of GEM tracking detectors will be inserted periodically just upstream
of the integrating detectors, at very low current, to track individual
particles during calibration runs.

\begin{figure}[ht]
\begin{center}
\includegraphics[width=3.25in]{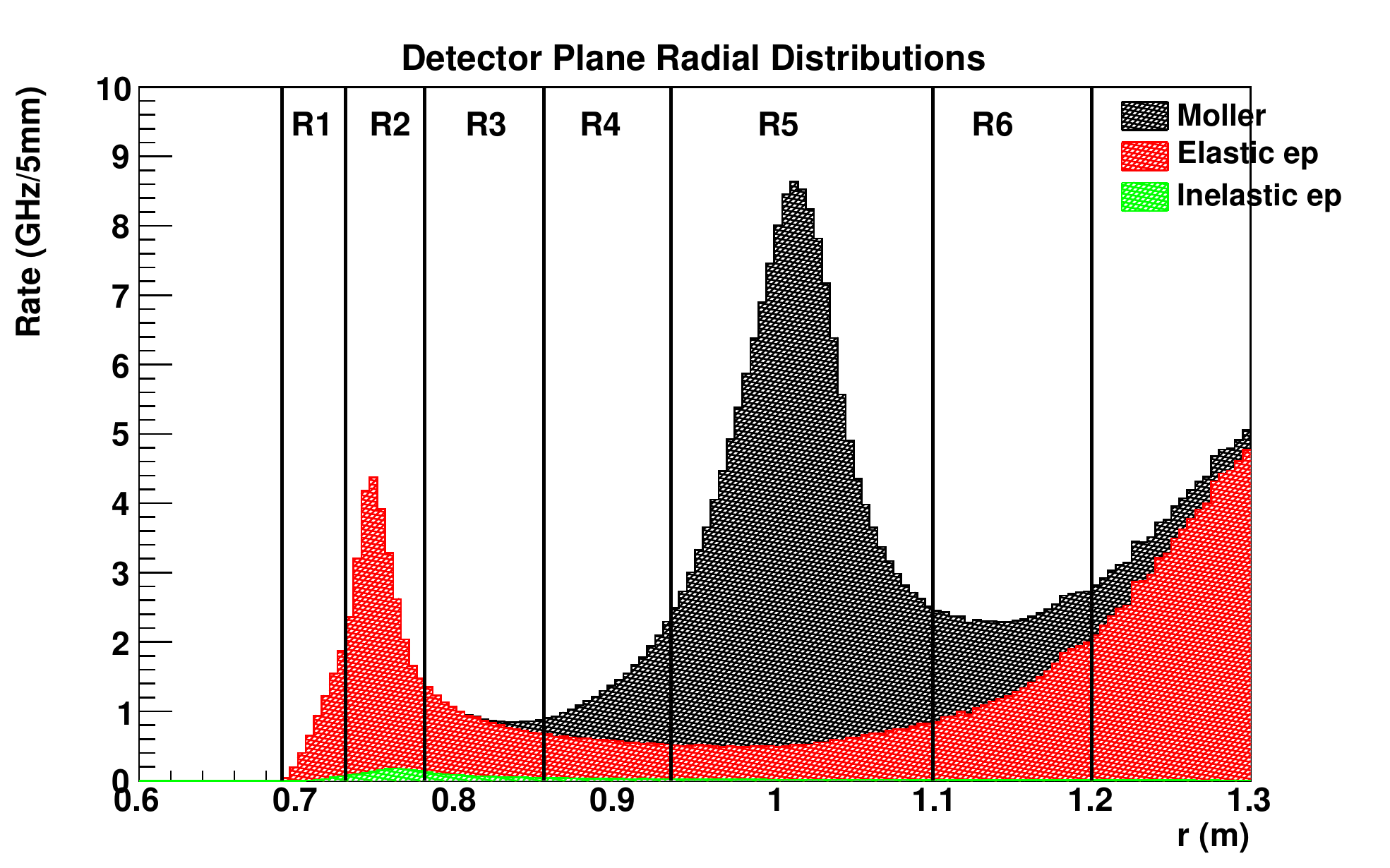}
\end{center}
\caption{Radial distribution of M\o ller (black), $ep$ elastic (red), and $ep$ inelastic electrons
28.5 m downstream of  target. The vertical black lines delineate the proposed radial segmentation
into 6 rings (R1 thru R6). The principal $A_{PV}$ measurement will be carried out in ring R5.}
\label{background_linear}
\end{figure}

The primary detectors are those labeled ``quartz assembly'' in Fig.~\ref{detectoroverview}.  Each detector is currently planned to consist of a piece of radiation hard fused silica (quartz) connected to a photomultiplier tube by a highly-reflective air-core light guide.  Prototype tests of detectors of this type have been done in a Mainz MAMI test beam with initial results ($> 25$ photoelectrons per event, $< 4 $\% excess noise) that already exceed the MOLLER specifications.  The focal plane will be segmented both radially (into 6 segments) and azimuthally (into 28 segments, with the critical M{\o}ller radial ring being more finely segmented into 84 segments) for a total of 224 total detector segments.   Simulations of the expected radial distribution of events at the focal plane are shown in Fig.~\ref{background_linear}.  The planned radial segmentation is indicated.  It is selected so that the primary M{\o}ller measurement will occur in Ring R5, while the other rings will allow the asymmetry of the backgrounds from elastic and inelastic electron-proton scattering to be simultaneously measured.

The MOLLER experiment constitutes a fourth generation of parity-violation experiment at Jefferson Lab.  It will benefit from the past experience and the trained user community while at the same time providing exciting challenges to attract and educate the next generation of experimental nuclear physicists.

\section{Conclusions}

The MOLLER experiment is designed to measure parity-violation in electron-electron scattering to unprecedented
precision using the 11 GeV electron beam in Hall A at JLab. The project represents a unique opportunity to 
probe physics beyond the Standard Model, with a unique window to new physics at MeV and multi-TeV scales.
The specific measurement described here would be the most sensitive low energy measurement of
a flavor-conserving purely leptonic interaction at low energy and cannot be carried out in any other existing or planned 
facility anywhere in the world. 
The project would realize its full potential towards the early part of the next decade, which is
timed well with anticipated results from high luminosity running at the LHC.
A motivated and experienced collaboration has been working on the design R\&D and 
is ready to carry out the full engineering design, construction, installation, data 
collection and analysis. 
Given the evolution of related projects, the timing of the 12 GeV upgrade, and the compelling
physics opportunity, the MOLLER project represents a compelling opportunity 
for investment during the period covered by the
2015 NSAC Long Range Plan.




\begin{acknowledgments}
This work has been supported by the United States Department of Energy, Office of Nuclear Physics, United States National Science Foundation, and the National Sciences and Engineering Research Council of Canada.

\end{acknowledgments}

\bibliographystyle{h-physrev3.bst}

\begin{thebibliography}{1}

\bibitem{MOLLER1}
MOLLER Experiment Homepage:
\begin{verbatim} http://http://hallaweb.jlab.org/12GeV/Moller/\end{verbatim}

\bibitem{MOLLER2}
MOLLER: Jefferson Lab Experiment E12-09-005:
\begin{verbatim} http://hallaweb.jlab.org/12GeV/Moller/pubs/moller_proposal.pdf\end{verbatim}

\bibitem{Anthony:2005pm}
  P.~L.~Anthony {\it et al.}  [SLAC E158 Collaboration],
  Phys.\ Rev.\ Lett.\  {\bf 95}, 081601 (2005)
  [arXiv:hep-ex/0504049].

\bibitem{Czarnecki:1995fw}
  A.~Czarnecki and W.~J.~Marciano,
  Phys.\ Rev.\  D {\bf 53}, 1066 (1996)
  [arXiv:hep-ph/9507420].
  
\bibitem{Czarnecki:2000ic}
  A.~Czarnecki and W.~J.~Marciano,
  Int.\ J.\ Mod.\ Phys.\  A {\bf 15}, 2365 (2000)
  [arXiv:hep-ph/0003049].


\bibitem{Erler:2004in}
  J.~Erler and M.~J.~Ramsey-Musolf,
  Phys.\ Rev.\  D {\bf 72}, 073003 (2005)
  [arXiv:hep-ph/0409169].

\bibitem{kkreview} 
  K.~S.~Kumar, S.~Mantry, W.~J.~Marciano and P.~A.~Souder,
  Ann.\ Rev.\ Nucl.\ Part.\ Sci.\  {\bf 63}, 237 (2013)
  [arXiv:1302.6263 [hep-ex]].
  
\bibitem{Cirigliano:2013lpa} 
  V.~Cirigliano and M.~J.~Ramsey-Musolf,
  Prog.\ Part.\ Nucl.\ Phys.\  {\bf 71}, 2 (2013)
  [arXiv:1304.0017 [hep-ph]].
  
\bibitem{Erler:2013xha} 
  J.~Erler and S.~Su,
  Prog.\ Part.\ Nucl.\ Phys.\  {\bf 71}, 119 (2013)
  [arXiv:1303.5522 [hep-ph]].

\bibitem{Chao:2014}
W.~Chao, H.~Li, M.~J.~Ramsey-Musolf, and S.~Su, in preparation.

\bibitem{Cirigliano:2004mv} 
  V.~Cirigliano, A.~Kurylov, M.~J.~Ramsey-Musolf and P.~Vogel,
  Phys.\ Rev.\ D {\bf 70}, 075007 (2004)
  [hep-ph/0404233].

\bibitem{Davoudiasl:2012ag} 
  H.~Davoudiasl, H.~-S.~Lee and W.~J.~Marciano,
  Phys.\ Rev.\ D {\bf 85}, 115019 (2012)
  [arXiv:1203.2947 [hep-ph]].
  
\bibitem{Davoudiasl:2012qa} 
  H.~Davoudiasl, H.~-S.~Lee and W.~J.~Marciano,
  Phys.\ Rev.\ Lett.\  {\bf 109}, 031802 (2012)
  [arXiv:1205.2709 [hep-ph]].

\bibitem{Kurylov:2003zh}
  A.~Kurylov, M.~J.~Ramsey-Musolf and S.~Su,
  Phys.\ Rev.\  D {\bf 68}, 035008 (2003)
  [arXiv:hep-ph/0303026].

\bibitem{RamseyMusolf:2006vr}
  M.~J.~Ramsey-Musolf and S.~Su,
  Phys.\ Rept.\  {\bf 456}, 1 (2008)
  [arXiv:hep-ph/0612057].

\bibitem{Erler:2011iw}
J.~Erler, P.~Langacker, S.~Munir and E.~Rojas,
arXiv:1108.0685v1 [hep-ph].

\bibitem{nsaclrp07}
  Nuclear Science Advisory Committee Long Range Planning Document (2007): 
 \begin{verbatim} http://hallaweb.jlab.org/12GeV/Moller/downloads/mie/nuclear_science_low_res.pdf\end{verbatim}

\bibitem{prescottreview} 
Report of the Director's Review (January 2010):
\begin{verbatim} http://hallaweb.jlab.org/12GeV/Moller/downloads/mie/
Final_Draft_MOLLER_Review_Report.pdf\end{verbatim}

\bibitem{nsacsubcomm12}
Report to NSAC on Implementing the 2007 Long Range Plan (2013):
\begin{verbatim} http://hallaweb.jlab.org/12GeV/Moller/downloads/mie/2013_NSAC_Implementing_the_2007_Long_Range_Plan.pdf\end{verbatim}

\bibitem{Amsler:2008zz}
  C.~Amsler {\it et al.}  [Particle Data Group],
  Phys.\ Lett.\  B {\bf 667}, 1 (2008).

\bibitem{csupdate}
  V.~A.~Dzuba, J.~C.~Berengut, V.~V.~Flambaum and B.~Roberts,
  Phys.\ Rev.\ Lett.\  {\bf 109}, 203003 (2012)
  [arXiv:1207.5864 [hep-ph]].

  
\bibitem{Zeller:2001hh}
  G.~P.~Zeller {\it et al.}  [NuTeV Collaboration],
  Phys.\ Rev.\ Lett.\  {\bf 88}, 091802 (2002)
  [Erratum-ibid.\  {\bf 90}, 239902 (2003)]
  [arXiv:hep-ex/0110059].

\bibitem{qweakfirst}
  D.~Androic {\it et al.}  [Qweak Collaboration],
  Phys.\ Rev.\ Lett.\  {\bf 111}, 141803 (2013)
  [arXiv:1307.5275 [nucl-ex]].
 

\bibitem{Souder:1990ia}
  P.~A.~Souder {\it et al.},
  Phys.\ Rev.\ Lett.\  {\bf 65}, 694 (1990).

\bibitem{Spayde:2003nr}
  D.~T.~Spayde {\it et al.}  [SAMPLE Collaboration],
  Phys.\ Lett.\  B {\bf 583}, 79 (2004)
  [arXiv:nucl-ex/0312016].

\bibitem{Acha:2006my}
  A.~Acha {\it et al.}  [HAPPEX collaboration],
  Phys.\ Rev.\ Lett.\  {\bf 98}, 032301 (2007)
  [arXiv:nucl-ex/0609002].

\bibitem{PREX}
The Lead Radius Experiment PREX, E06002, K. Kumar, R. Michaels, P. Souder, G. Urciuoli spokespersons,
http://hallaweb.jlab.org/parity/prex/ .

\bibitem{Qweak}
The Qweak Experiment, http://www.jlab.org/Hall-C/Qweak/index.html, , R. Carlini Principal Investigator.

\bibitem{Smith:2012}
G.~R.~Smith, 
Il Nuovo Cimento C {\bf 035N4} 159 (2012).

\bibitem{Derman:1979zc}
  E.~Derman and W.~J.~Marciano,
  Annals Phys.\  {\bf 121}, 147 (1979).

\bibitem{n34} 
  A.~Aleksejevs, S.~Barkanova, Y.~Kolomensky, E.~Kuraev and V.~Zykunov,
  Phys.\ Rev.\ D {\bf 85}, 013007 (2012)
  [arXiv:1110.1750 [hep-ph]].

\bibitem{n35} 
  A.~G.~Aleksejevs, S.~G.~Barkanova, Y.~M.~Bystritskiy, A.~N.~Ilyichev, E.~A.~Kuraev and V.~A.~Zykunov,
  Eur.\ Phys.\ J.\ C {\bf 72}, 2249 (2012).

\bibitem{n36}
  A.~G.~Aleksejevs, S.~G.~Barkanova, V.~A.~Zykunov and E.~A.~Kuraev,
  Phys.\ Atom.\ Nucl.\  {\bf 76}, 888 (2013)
  [Yad.\ Fiz.\  {\bf 76}, 942 (2013)].

\bibitem{csapv}
  S.~C.~Bennett and C.~E.~Wieman,
  Phys.\ Rev.\ Lett.\  {\bf 82}, 2484 (1999)
  [Erratum-ibid.\  {\bf 83}, 889 (1999)]
  [hep-ex/9903022].

\bibitem{:2005ema}
    [ALEPH Collaboration and DELPHI Collaboration and L3 Collaboration and
    OPAL Collaboration and SLD Collaboration and LEP Electroweak Working Group and SLD Electroweak
     Group and SLD Heavy Flavour Group ],
  Phys.\ Rept.\  {\bf 427}, 257 (2006)
  [arXiv:hep-ex/0509008].
 
\bibitem{Marciano:2006zu}
  W.~J.~Marciano,
  AIP Conf.\ Proc.\  {\bf 870}, 236 (2006).

\bibitem{Eichten:1983hw}
  E.~Eichten, K.~D.~Lane and M.~E.~Peskin,
  Phys.\ Rev.\ Lett.\  {\bf 50}, 811 (1983).

\bibitem{newjenszprime}
J.~Erler, P.~Langacker, S.~Munir and E.~Rojas,
arXiv:1108.0685v1 [hep-ph].

\bibitem{petriello}
  Y.~Li, F.~Petriello and S.~Quackenbush,
  Phys.\ Rev.\  D {\bf 80}, 055018 (2009)
  [arXiv:0906.4132 [hep-ph]].
    
\bibitem{Davoudiasl:2014kua} 
  H.~Davoudiasl, H.~-S.~Lee and W.~J.~Marciano,
  arXiv:1402.3620 [hep-ph].
  
\bibitem{snomass:2013}
M. Baak, {\it et al.}, 
Snomass 2013 study,
arXiv:1310.6708v1 [hep-ph].

\bibitem{nusong:2009}
 T.~Adams, {\it et al.},
 Int. J. Mod. Phys. A {\bf 26}, 671 (2009).

\bibitem{Conrad:2005}
  J.~M.~Conrad, J.~M.~Link, and M.~H.~Shaevitz,
  Phys.\ Rev.\ D {\bf 71}, 073013 (2005)
  [arXiv:hep-ex/0403048].

\end{thebibliography}

\pagebreak

\appendix
\section{MOLLER Collaboration List}
 
 \begin{center}
\begin{large}
The MOLLER Collaboration \\
\end{large}

\begin{small}
\vspace*{3.0ex}
J. Benesch, P. Brindza, R.D. Carlini, J-P.~Chen, E.~Chudakov, S.~Covrig, M.M.~Dalton,
A.~Deur, D.~Gaskell, A.~Gavalya,
J.~Gomez, D.W.~Higinbotham, C.~Keppel, D.~Meekins, R.~Michaels, B.~Moffit, Y.~Roblin, R.~Suleiman,  R.~Wines,
B.~Wojtsekhowski \\
    {\em Jefferson Lab}
% ...
    \vspace*{0.9ex} \\
 $^*$kkumar@physics.umass.edu     \\ 

\end{small}
\end{center}

\end{document}